\documentclass[format=acmsmall, review=false,nonacm]{acmart}
\pdfoutput=1
\usepackage{acm-ec-25-proc}
\usepackage{booktabs} 
\usepackage[ruled]{algorithm2e} 
\usepackage{natbib} 

\SetAlFnt{\small}
\SetAlCapFnt{\small}
\SetAlCapNameFnt{\small}
\SetAlCapHSkip{0pt}
\IncMargin{-\parindent}

\setcitestyle{authoryear}

\ccsdesc[500]{Mathematics of computing~Stochastic processes}

\keywords{Stochastic Games, Strategy Complexity, B\"uchi, Transience}

\copyrightyear{2026}
\acmYear{2026}
\setcopyright{cc}
\setcctype{by}
\acmConference[Full version of a paper presented at EC '25]{}{}{}
\acmBooktitle{}
\acmDOI{}
\acmISBN{}

\input{preamble}
\input{macro}
\usetikzlibrary{arrows,decorations,intersections,pgfplots.fillbetween}
\usetikzlibrary{shapes.multipart,shapes.misc}
\usepgflibrary{decorations.pathreplacing}
\usetikzlibrary{arrows,topaths}
\usetikzlibrary{automata,shapes.arrows, backgrounds, fadings,intersections, positioning}
\usetikzlibrary{shadows}

\usetikzlibrary{decorations.pathmorphing}
\usetikzlibrary{decorations.shapes}
\usetikzlibrary{shapes}
\usetikzlibrary{backgrounds}
\usetikzlibrary{fit}
\usetikzlibrary{arrows}
\usetikzlibrary{calc}
\usetikzlibrary{automata}
\usetikzlibrary{shapes.geometric}

\tikzset{every picture/.style={thick,>=angle 60, node distance=2cm and 2cm}}
\tikzset{Grand/.style={draw,circle,minimum size=11*1.5,inner sep=0}}
\tikzset{Gmax/.style={draw,rectangle,minimum size=9*1.5,inner sep=0}}
\tikzset{Gmin/.style={draw,diamond,minimum size=9*1.5,inner sep=0}}
\tikzset{gamebad/.style={fill=red}}

\tikzset{every loop/.style={looseness=20}}
\tikzset{>=stealth, shorten >=1pt, shorten <=1pt}

\tikzset{state/.style={draw,circle,inner sep=2,minimum size=17pt}}

\tikzset{square/.style={regular polygon,regular polygon sides=4}}
\tikzset{acta/.style={
fill=blue!40!white,square,inner sep=2,
}}
\tikzset{actb/.style={
fill=red!40!white,diamond,inner sep=1,
}}

\tikzset{elliptic state/.style={draw,ellipse, minimum width=30, minimum height=15}}

\AtEndPreamble{%
  \theoremstyle{acmplain}
  \newtheorem{remark}[theorem]{Remark}
  
}

\title[Strategy Complexity of B{\"u}chi Games]{Strategy Complexity of B{\"u}chi and Transience Objectives in Concurrent Stochastic Games}

\author{Stefan Kiefer}
\orcid{0000-0003-4173-6877}
\affiliation{%
  \institution{University of Oxford}
  \city{Oxford}
  \country{United Kingdom}
}

\author{Richard Mayr}
\orcid{0009-0003-2164-9933}
\affiliation{%
  \institution{University of Edinburgh}
  \city{Edinburgh}
  \country{United Kingdom}
}

\author{Mahsa Shirmohammadi}
\orcid{0000-0002-7779-2339}
\affiliation{%
  \institution{IRIF \& CNRS, Universit\'e Paris cit\'e}
  \city{Paris}
  \country{France}
}

\author{Patrick Totzke}
\orcid{0000-0001-5274-8190}
\affiliation{%
 \institution{University of Liverpool}
  \city{Liverpool}
  \country{United Kingdom}
}

\begin{abstract}
We study 2-player stochastic games on countable graphs. 
Players Max and Min seek respectively to maximize and minimize the
probability of satisfying the game objective.
The B\"uchi objective is to visit a given set of states infinitely
often.
The Transience objective is to visit no state infinitely often.
In B\"uchi games there exist $\eps$-optimal Max strategies that use just a
step counter plus 1 bit of public memory. This upper bound holds for all
countable graphs, but is a new result even for finite graphs.
It is tight, since Max strategies that use just a step
counter, or just finite memory, are \emph{not} sufficient even on finite game graphs.
This upper bound follows from a slightly stronger new result:
$\eps$-optimal Max strategies for the \emph{combined} B\"uchi and Transience objective
require exactly 1 bit of public memory.
Moreover, $\eps$-optimal Max strategies for the Transience objective alone
can be chosen as memoryless.

\end{abstract}

\begin{document}

\maketitle

\section{Introduction}
\label{sec-intro}
\mysec{Background}
We study 2-player zero-sum concurrent (i.e., simultaneous move) stochastic games on
countable (finite or countably infinite) graphs.
\footnote{Our proofs do not carry over to uncountable state spaces.
Apart from the usual issues with measurability in uncountable systems,
we also, e.g., partition events into as many parts as there are states
(in our case only countably many) and then rely on sigma-additivity.}
Introduced by Shapley in his seminal 1953 work~\cite{shapley1953}, 
and generalized in \cite{Gillette1958} and \cite{KumarShiau} 
to allow infinite state and action sets and non-termination,
such games play a central role in the solution of many problems
in
Evolutionary Biology \cite{raghaven2012}, 
Computer Science \cite{BordaisB022,bordais_et_al:LIPIcs.FSTTCS.2022.33,AlfaroH01,neyman2003,AltmanAMM05, Altman2007,solan2015stochastic,svorevnova2016,bouyer2016},
Economics \cite{sorin1992,nowak2005,jaskiewiczN11,solan2015stochastic,bacharach2019},
and Operations Research
\cite{doi:10.1287/moor.1080.0372,doi:10.1287/moor.2021.1123,doi:10.1287/moor.26.2.265.10549,doi:10.1287/moor.7.1.14,Osmani2026,KMST:AOR2026}.

General objectives are defined via measurable reward functions,
where the two players, Max and Min, aim to maximize (resp.\ minimize)
the reward of Max.
For the subclass of event-based objectives,
the reward function is the indicator function of some event
(a measurable set of plays), i.e., the players aim to maximize
(resp.\ minimize) the probability of this event.

The \emph{B\"uchi objective} is an event-based objective where
Max aims to visit a given subset of the states
infinitely often.
It can be understood as a special case of the objective to maximize
the expected $\limsup$ of the infinite sequence of daily rewards in a play.
If all daily rewards are in $\{0,1\}$ then Max needs to maximize the
probability of visiting infinitely often the subset of the states with reward $1$.
(The term \emph{B\"uchi objective} originates in automata theory, e.g., \cite{GTW:2002}.)

The \emph{Transience} objective is to visit no state infinitely
often, i.e., every single state is visited at most finitely often (if at all).
This is equivalent to the condition that every \emph{finite} subset of the
states is eventually left forever.
It can be interpreted as a notion of perpetual progress (pushing outward).

A central result in zero-sum 2-player stochastic games with finite action sets 
is the existence of a \emph{value} for the large class of Borel measurable
objectives \cite[]{martin_1998,Maitra-Sudderth:1998}.
That is, for Borel measurable functions $f$ on plays,
\[
\sup_{\it Max}\inf_{\it Min} f = {\it value} = \inf_{\it Min}\sup_{\it Max} f
\]
over Max and Min strategies.
This holds even under the weaker assumption that
one player has countable action sets and the other plays has finite action sets
\cite{martin_1998}\cite[Theorem 11]{Flesch-Predtetchinski-Sudderth:2020}.
Throughout this paper we assume that Max has countable action sets and Min has
finite action sets.
(While the reverse case also has value, our results do not hold for it; see below.)

The amount/type of memory and randomization required for an
$\eps$-optimal strategy for a
given player and objective is also called the \emph{strategy complexity}.
One can also consider the strategy complexity of optimal strategies
in cases where optimal strategies exist.

Optimal Max strategies for B\"uchi objectives do not exist in general,
not even for the simpler reachability objective in finite-state concurrent games
\cite{AlfaroHK98}.
However, in finite-state concurrent B\"uchi games, if every state admits an
optimal Max strategy then
Max also has a memoryless strategy that is optimal from all states
\cite[Proposition 17]{bordais_et_al:LIPIcs.FSTTCS.2022.33}.
Optimal Max strategies for Transience do not exist even in countable
Markov decision processes (MDPs)
\cite{KMST:CONCUR2021}, and Transience trivially never holds in finite-state MDPs/games.
Thus we consider the strategy complexity of $\eps$-optimal strategies.

\mysec{Our contribution}
We show (\cref{thm:Buchi}) that $\eps$-optimal Max strategies for the B\"uchi objective in countable
concurrent stochastic games require just a step
counter plus 1 bit of public memory.
This is a new result even for the subclass of \emph{finite-state} concurrent
stochastic games with finitely many actions,
where the strategy complexity of B\"uchi objectives had also been open.
(About the strategy complexity for the opposing Min player see \Cref{sec-minimizer}.)

The step counter plus 1 bit upper bound on the strategy complexity
of $\eps$-optimal Max strategies is tight, even for finite-state concurrent
B\"uchi games.
In the ``Bad Match'' game \cite[Example 13.7]{MaitraSudderth:DiscreteGambling},
both finite-memory strategies and Markov strategies (i.e., strategies that use just a step counter)
are worthless, i.e.,
they are not $\eps$-optimal for Max for any $\eps < 1$; see \Cref{sec-bad-match}. 
(The fact that finite-memory Max strategies are worthless in finite-state
concurrent B\"uchi games was mentioned (without proof)
in \cite{Alfaro-Henzinger:LICS2000}, who considered a different
counterexample, namely a repeated version of
the classic ``Snowball game'' (aka Hide-or-Run game) of
\cite[Example 1]{Everett1957} and \cite{KumarShiau}.)
In the special case of finite-state \emph{turn-based}
(aka alternating move)
B\"uchi games, the strategy
complexity is lower, because both players have optimal memoryless
deterministic strategies.
For countable MDPs, different counterexamples show that
neither finite-memory strategies nor Markov strategies can be 
$\eps$-optimal for Max for the B\"uchi objective
for any $\eps < 1$ \cite{KMST:ICALP2019}.

Our upper bound (\cref{thm:Buchi}) is a consequence of a stronger new
result (\cref{thm:TrBuchi}) on the strategy complexity of the \emph{combined} B\"uchi and
Transience objective in countable concurrent stochastic games with finite Min
action sets:
$\eps$-optimal Max strategies require just 1 bit of (public) memory.
This upper bound is tight, even for acyclic countable MDPs,
in the sense that memoryless Max strategies do not suffice \cite{KMST:ICALP2019}.
(Note that plays can only satisfy this combined objective if the 
target set of the B\"uchi objective is infinite).
The stronger result on the combined B\"uchi and
Transience objective implies the result on the B\"uchi
objective above as follows.
We can apply the stronger result to a new derived acyclic game
where the step counter is encoded into the states (and thus Transience always
holds), and carry the obtained
1-bit Max strategy back to the original game as a step counter plus 1 bit
strategy; see \Cref{prop:redacyclic}.

Our upper bound on the strategy complexity of the combined B\"uchi and
Transience objective in countable concurrent stochastic games
generalizes a corresponding result for countable MDPs \cite[Lemma 4]{KMST:CONCUR2021}.
Similarly to \cite{KMST:CONCUR2021}, 1 bit of public memory is used to
switch between two modes. Roughly speaking,
one mode aims to visit the target set in the short term
while the other mode records that this short-term 
goal has been attained.
However, the proof techniques for stochastic games are very different from
those used for MDPs. Unlike in MDPs, Max cannot statically plan ahead which
states in the B\"uchi target set will be visited within a given time horizon with high
probability, because of the influence of Min's decisions.
Similarly, for transience, Max cannot statically plan that certain states
are very unlikely to be visited again after a given time horizon.
Again it may depend on Min's decisions whether certain states are (re-)visited
arbitrarily late in the game.
Therefore, our construction of the Max strategy is based on a different
principle.
It proceeds state by state (based on an enumeration of the states),
rather than phase by phase as was done for MDPs in \cite{KMST:CONCUR2021}.

Our third result (\cref{thm:Tr}) concerns the strategy complexity of the Transience objective
alone (which is only meaningful in infinite-state games).
There always exist memoryless $\eps$-optimal Max strategies for Transience in
countable concurrent stochastic games.
This generalizes a corresponding result for countable MDPs
\cite[Theorem 8]{KMST:CONCUR2021}. However, unlike for MDPs, these strategies cannot be
deterministic in concurrent games, i.e., they need to randomize.

Finally, our upper bounds do not carry over to countable stochastic
games with infinite Min action sets.
Even for the simpler reachability objective, there is 
a counterexample where Max strategies that just use a
step counter plus finite (private) memory are not $\eps$-optimal for any
$\eps < 1$. In fact, the counterexample is a 
countably infinite-state \emph{turn-based} stochastic reachability
game with countably infinite Min branching and finite Max branching \cite{KMSTW:DGA}.

\section{Preliminaries}
\label{sec-prelim}
A \emph{probability distribution} over a countable set $\states$ is a function
$\alpha:\states\to[0,1]$ with $\sum_{\state\in \states}\alpha(\state)=1$.
The \emph{support} of~$\alpha$ is the set $\{\state \in \states \mid \alpha(\state) >0\}$.
$\dist(\states)$ is the set of all probability distributions over~$\states$.
\mysec{Stochastic Games}
We study stochastic games between two players, Max and Min.
A \emph{(concurrent) game}~$\game$ is played on a countable set of states~$\states$.
For each state $\state \in \states$ there are non-empty countable \emph{action} sets
$A(\state)$ and $B(\state)$ for Max and Min, respectively.
A \emph{mixed action} for Max (resp.\ Min) in state $\state$
is a distribution over $A(\state)$ (resp.\ $B(\state)$).
Let $A\eqdef\bigcup_{s\in\states}A(s)$ and
$B\eqdef\bigcup_{s\in\states}B(s)$.

Let $Z \eqdef \{(\state,a,b) \mid \state \in \states,\ a \in A(\state),\ b \in B(\state)\}$.
For every triple $(\state,a,b) \in Z$
there is a distribution $p(\state,a,b) \in \dist(\states)$ over successor states.
We call a state~$s \in S$ a \emph{sink} state if
$p(s,a,b)(s) = 1$ for all $a \in A(s)$ and $b \in B(s)$.
We extend the \emph{transition function}~$p$ to mixed actions $\alpha \in \dist(A(s))$ and $\beta \in \dist(B(s))$ by letting
\[
p(s,\alpha,\beta)\,\,\eqdef \sum_{a \in A(s)} \sum_{b \in B(s)} \alpha(a) \beta(b) p(s,a,b),\] which is a distribution over~$S$.
Let $\state_0\in \states$ be an initial state. 
A \emph{play} from~$\state_0$ is an infinite
sequence $(s_0,a_0,b_0)(s_1,a_1,b_1)\dots$ in $Z^\omega$ where
$s_{n+1}$ is in the support of $p(s_n,a_n,b_n)$.
The game is played in steps,
where at every step $n \in \N$ the play is in some state~$\state_n$,
and Max and Min independently and simultaneously choose mixed actions
$\alpha_n \in \dist(A(\state_n))$ and $\beta_n \in \dist(B(\state_n))$.
Then actions $a_n\in A$ and $b_n\in B$ are sampled according to $\alpha_n$ and
$\beta_n$, respectively, and the state $\state_{n+1}\in\states$ is
chosen according to $\probp(\state_n,a_n,b_n)$,
before both players are informed of the outcomes
$a_{n},b_{n},s_{n+1}$ and the next step starts from $\state_{n+1}$.
The game is called \emph{turn-based} (or alternating move) if every
$\probp(\state,a,b)$ either solely depends on $(s,a)$ or solely on $(s,b)$,
i.e.,
for each $s$,
only the action of one player matters.

\mysec{Strategies}
We write $H_n$ for the set of \emph{histories} at step $n\in\N$, which is defined
as
$H_0 \eqdef \states$ and $H_n \eqdef Z^n \times \states$ for all $n>0$.
The set of all histories
is $H \eqdef \bigcup_{n \in \N} H_n$.
For each history $h = (s_0,a_0,b_0) \cdots (s_{n-1},a_{n-1},b_{n-1}) s_{n} \in H_n$, 
let $\state_h \eqdef s_{n}$ denote its final state.
Define the \emph{length} of~$h$ by $\len{h} \eqdef n$.
For any history
$h'$ starting in state $s_h$,
we write $hh'$
for the concatenation of the histories
$(s_0,a_0,b_0) \cdots(s_{n-1},a_{n-1},b_{n-1}) h'$.
A \emph{strategy} for Max
is a function $\zstrat$ that to each history~$h \in H$ assigns
a mixed action $\zstrat(h) \in \dist(A(\state_h))$.
It is called \emph{deterministic} if $\zstrat(h)$ is a Dirac distribution for every history $h$.
Mixed actions and strategies $\ostrat$ for Min are defined analogously.
We write $\zstratset$ 
and $\ostratset$ for the sets of strategies for Max and Min, respectively.

\mysec{Probability Measures and Optimality} 
An initial state $\state_0$ and a pair of strategies $\zstrat, \ostrat$
for Max and Min induce a probability measure
where the events, also called \emph{objectives} here, are
measurable sets of plays
(see e.g.\ \cite{Flesch-Predtetchinski-Sudderth:2020}).
We write $\probm_{\game,\state_0,\zstrat,\ostrat}({\playset})$ for the
probability of the event $\playset$ starting from~$\state_0$.
(In the special case of MDPs with only one active player,
we only denote the strategy of that player, $\zstrat$ or $\ostrat$, resp.)
It is initially defined for the cylinder sets generated by the histories
and then extended to the sigma-algebra by Carath\'eodory's
unique extension theorem~\cite{billingsley-1995-probability}.
Given some Borel measurable reward function $v$ on plays,
we will write $\expectval_{\game,\state_0,\zstrat,\ostrat}$
for the expectation w.r.t.~$\probm_{\game,\state_0,\zstrat,\ostrat}$ and $v$.
We may drop~$\game$ from the subscript when it is understood.
For a game $\game$ and objective $\playset$ fixed, 
the \emph{lower value} and \emph{upper value} of a state~$\state_0$ are defined as
\[
\valueoflower{\game,\playset}{\state_0} 
    \eqdef \sup_{\zstrat \in \zstratset} \inf_{\ostrat \in \ostratset}
    \probm_{\game,\state_0,\zstrat,\ostrat}({\playset}) \quad\text{and}\quad
\valueofupper{\game,\playset}{\state_0} 
   \eqdef \inf_{\ostrat \in \ostratset} \sup_{\zstrat \in \zstratset}
\probm_{\game,\state_0,\zstrat,\ostrat}({\playset}).
\]
The inequality $\valueoflower{\game,\playset}{\state_0} \le \valueofupper{\game,\playset}{\state_0}$ holds by definition.
If $\valueoflower{\game,\playset}{\state_0} = \valueofupper{\game,\playset}{\state_0}$, then this quantity is called the
\emph{value} of $\state_0$,
denoted by $\valueof{\game,\playset}{\state_0}$.
For all Borel objectives, the value exists if all action sets are finite,
and even if, for all states~$s$, we have that $A(s)$ is finite or $B(s)$ is
finite~\cite{martin_1998}\cite[Theorem 11]{Flesch-Predtetchinski-Sudderth:2020}.
We always assume the latter, 
so that $\valueof{\game,\playset}{s_0}$ exists.

For $\eps \ge 0$, a Max strategy~$\zstrat$ is called \emph{$\eps$-optimal} from~$s_0$ if for all Min strategies~$\pi$ we have
\(\probm_{\game,\state_0,\zstrat,\ostrat}({\playset})\ge {\valueof{\game,\playset}{\state_0} - \eps}.\)
It is called \emph{multiplicatively $\eps$-optimal} if
it satisfies
\(\probm_{\game,\state_0,\zstrat,\ostrat}({\playset})\ge \valueof{\game,\playset}{\state_0} (1-\eps).\)
Clearly, $\zstrat$ is $\eps$-optimal if it is multiplicatively $\eps$-optimal.
A $0$-optimal strategy is also called \emph{optimal}.

\mysec{Objectives}
Given a set $\reachset \subseteq \states$ of target states, the \emph{reachability} objective $\reach[\reachset]$ is the set of plays that visit $\reachset$ at least once, i.e., $s_h \in \reachset$ holds for some history~$h$ that is a prefix of the play.
For $Q \subseteq \reachset$, the objective $\reachn{Q}{\reachset}$ additionally requires that all states in
$h$ \emph{before} the visit to $T$ are in~$Q$.
The dual \emph{safety} objective~$\avoid[\reachset] \eqdef Z^\omega \setminus \reach[\reachset]$ consists of the plays that never visit~$T$.
We are mainly interested in \emph{B\"uchi} objectives $\Buchi[T]$
that consist of all those plays that visit the set of target states
$T \subseteq S$ infinitely often.
(If the set $T$ is infinite then some plays that satisfy $\Buchi[T]$
might not visit any particular state in $T$ infinitely often.)
Whenever possible, we will assume a fixed target set~$T$ and simply write $\reach,\avoid$ and $\Buchi$ for these objectives.
The \emph{transience}  objective $\Tr$ contains those plays in which every state of the game 
is eventually avoided.
That is, a play 
$(s_0,a_0,b_0)(s_1,a_1,b_1)\cdots \in Z^\omega$
is in 
$\Tr$ if for every $\state\in\states$ there exist $n_\state$ so that $\state\neq s_i$ for $i\ge n_\state$.
The \emph{transient B\"uchi} objective is $\TB \eqdef \transient \cap \Bu$.
Clearly, $\TB$~can only be met if $\reachset$~is infinite.
An event $E \subseteq Z^\omega$ is called \emph{shift-invariant} iff
$(s_0,a_0,b_0)(s_1,a_1,b_1)\cdots \in E \ \Leftrightarrow\
(s_i,a_i,b_i)(s_{i+1},a_{i+1},b_{i+1})\cdots \in E$ for any $i\in \N$,
i.e., a finite prefix is removed.
Note that the events $\Bu, \transient, \TB$ are all shift-invariant.

%

\mysec{Strategy Classes}
We formalize the amount of \emph{memory} needed to implement strategies.
%
A Max (analogously Min) strategy is \emph{based on} a set $M$
of \emph{memory modes} if it can be described by an
initial mode $m_0\in M$ and a
pair of functions
$\stratAct  : \states \x M \to \dist(A)$ and 
$\stratUp : \states\x M \x A\x B\x \states \to \dist(M)$,
{selecting} mixed actions and
{updating} the memory mode, respectively.
For any history
$h = (s_0,a_0,b_0) \cdots (s_{n-1},a_{n-1},b_{n-1}) s_{n} \in H_n$,
$\sigma(h) = \stratAct(s_n,m_n)$ and
$m_{n+1}$ is drawn randomly according to 
{the distribution}
\(\stratUp(s_n,m_n,a_{n},b_{n},s_{n+1}).\)
For $M$-based strategy $\zstrat$ and $m\in M$ we will write $\zstrat[m]$ to denote the strategy with $\stratAct$ and $\stratUp$ as in $\zstrat$ but with initial mode~$m_0=m$.

A memory-based strategy is \emph{public} if at each step of the game, its current memory mode is known to the opponent,
and \emph{private} otherwise.
Note that if all memory updates $\stratUp(.)$ are Dirac then $\sigma$ is
public, since the opponent can keep track of its memory.
%
A strategy $\zstrat$ is \emph{memoryless}
if it is based on a singleton set~$M$, i.e., $\zstrat(h)=\zstrat(h')$ for any two histories $h,h'$ with $\state_h=\state_{h'}$.
A strategy is \emph{1-bit} iff
$M=\{0,1\}$.
A \emph{step counter} is a special case of infinite memory in the form of a discrete
clock (by default starting as $0$) that gets incremented by $1$ at every step, independently of the actions of the
players. Strategies that use just a step counter are also called \emph{Markov} strategies.
A strategy $\zstrat$ is \emph{1-bit Markov} if it is based on a step counter
and two extra memory modes, i.e., $M=\N\x\{0,1\}$
and at any step $n$ the update function $\stratUp$ 
guarantees that the mode $m_n \in \{n\}\x\{0,1\}$.

\section{The Main Results}
\label{sec-overview}
Our results concern the memory requirement of Max strategies for
the B\"uchi ($\Buchi$), Transience ($\transient$) and Transient B\"uchi ($\TB$) objectives in concurrent games.

Throughout the paper, $\game$ refers to a concurrent 2-player game with a countable state space $\states$,
where Max has countable action sets and Min has finite action sets.

\begin{restatable}{theorem}{thmBuchi}\label{thm:Buchi}
Let $\game$ be a game with B\"uchi objective and $\eps>0$. 
Max has a 1-bit Markov strategy $\zstrat$ such that
\begin{enumerate}
\item $\zstrat[0]$ is multiplicatively $\eps$-optimal from every state,
  \footnote{Recall that $\zstrat[0]$ means that the 1-bit memory is initially
    set to $0$. In contrast, the initial value of the step counter does not matter.
    The strategy is multiplicatively $\eps$-optimal from every state,
    regardless of the initial value of the step counter, and
thus in particular in the default case where the step counter is initially $0$.}
  and
    \item all memory updates $\stratUp(\cdot)$ are Dirac (hence the memory is public).
\end{enumerate}
Moreover, if $\game$ is turn-based then there exists such a strategy that is deterministic. 
\end{restatable}

The 1-bit Markov upper bound
for the B\"uchi objective of \Cref{thm:Buchi} is a new result
even for the special case of \emph{finite-state} concurrent B\"uchi games.
Moreover, this upper bound is tight in the sense that Markov strategies or
finite-memory strategies are not sufficient, neither for finite-state
concurrent games \cite[Example 13.7]{MaitraSudderth:DiscreteGambling},
\cite{Alfaro-Henzinger:LICS2000} nor for countable MDPs \cite{KMST:ICALP2019}. 
We re-visit some of these lower bounds in
\cref{sec-lower-bounds} where we discuss in detail the example of the ``Bad Match'',
in a slightly adapted formulation as a finite-state concurrent B\"uchi game.

Our strongest result is that for transient B\"uchi objectives:
Max has public 1-bit strategies that are uniformly $\eps$-optimal
(i.e., $\eps$-optimal from every state).

\begin{restatable}{theorem}{thmTrBuchi}\label{thm:TrBuchi}
Let $\game$ be a game with $\TB$ objective and $\eps>0$.
Max has a 1-bit strategy $\zstrat$
such that 
\begin{enumerate}
    \item $\zstrat[0]$ is multiplicatively $\eps$-optimal from every state, and
    \item all memory updates $\stratUp(\cdot)$ are Dirac (hence the memory is public).
\end{enumerate}
Moreover, 
if $\game$ is turn-based then there exists such a strategy that is deterministic.
\end{restatable}

This implies \cref{thm:Buchi}.
Indeed, following~\cite{KMST:CONCUR2021}, 
for any game $\game$ on $\states$ we can consider the game on the product $\states\x\N$
in which the second component acts as the step counter.
The resulting game $\game'$ is acyclic and therefore all plays are transient.
By applying \Cref{thm:TrBuchi} to $\game'$, we obtain an $\eps$-optimal 1-bit
Max strategy, which can be carried back to $\game$ as a 1-bit Markov strategy.
See \cref{sub:proof-thm-Buchi} for more details.  

\Cref{tab:summary} summarizes known results about the strategy complexity of
B\"uchi objectives.
\begin{table}
{\small
\begin{tabular}{lcc}
& one (or finitely many) target states &
                                                                             arbitrary
                                                                             set
                                                                             of
                                                                             target states\\ \hline
Finite-state  & \multicolumn{2}{|c|}{Memoryless deterministic \cite{Zielonka04}}\\
turn-based game & \multicolumn{2}{|c|}{}\\ \hline
Finite-state                 & \multicolumn{2}{|c|}{Randomized 1-bit Markov (\cref{thm:Buchi})}\\
concurrent game              & \multicolumn{2}{|c|}{Not finite-memory (\cref{claim:badmatch-finite}) }\\  
                             & \multicolumn{2}{|c|}{Not Markov (\cref{prop:bad-match-no-markov}) }\\ \hline
Infinite-state               & \multicolumn{1}{|c|}{Deterministic 1-bit Markov (\cref{thm:Buchi})} & \multicolumn{1}{|c|}{Deterministic 1-bit Markov (\cref{thm:Buchi})} \\ 
MPD                          & \multicolumn{1}{|c|}{Not finite-memory \cite{KMST:AOR2026}}  &\multicolumn{1}{|c|}{ Not finite-memory \cite{KMST:AOR2026}}\\
                             & \multicolumn{1}{|c|}{{\bf Open} if Markov}        & \multicolumn{1}{|c|}{Not Markov \cite{KMST:AOR2026}}\\ \hline
Infinite-state               & \multicolumn{2}{|c|}{Deterministic 1-bit Markov (\cref{thm:Buchi})} \\ 
turn-based game              & \multicolumn{2}{|c|}{Not finite-memory \cite{KMST:AOR2026}  }\\
                             & \multicolumn{2}{|c|}{Not Markov (\cref{prop:bad-match-no-markov-turnbased}) }\\ \hline
Infinite-state               & \multicolumn{2}{|c|}{Randomized 1-bit Markov (\cref{thm:Buchi})}\\
concurrent game              & \multicolumn{2}{|c|}{Not finite-memory (\cref{claim:badmatch-finite}) }\\
                               & \multicolumn{2}{|c|}{Not Markov (\cref{prop:bad-match-no-markov})} \\ \hline
\end{tabular}
}
\caption{Summary of the strategy complexity of $\eps$-optimal strategies for
  B\"uchi objectives in MDPs and stochastic games.}\label{tab:summary}
\end{table}

Our third main contribution is
that in games with Transience objectives, Max has memoryless strategies that are uniformly $\eps$-optimal.

\begin{restatable}{theorem}{thmTr}\label{thm:Tr}
Let $\game$ be a game with $\transient$ objective and ${\eps>0}$.
Max has a memoryless strategy that is multiplicatively
$\eps$-optimal from every state.

Moreover, if $\game$ is turn-based then there exists such a strategy that is deterministic.
\end{restatable}

The technique we use to prove this result is similar to, but simpler than the one used for \cref{thm:TrBuchi}.
We therefore first focus on transience games in \Cref{sec-trans}, before proving \cref{thm:TrBuchi,thm:Buchi} in \cref{sec-buchi}.

\section{Lower Bounds}
\label{sec-lower-bounds}
\label{sec-bad-match}
\begin{figure}[t]
  \centering
\begin{minipage}{.47\textwidth}
  \centering
    \includegraphics{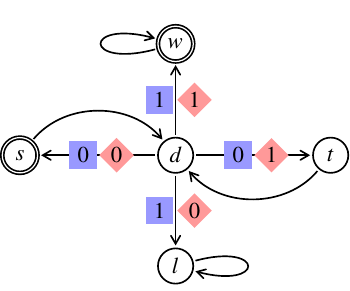}
\captionof{figure}{The Bad Match.
Max and Min actions, where they influence the motion, are depicted in blue and red, respectively.
The objective is $\Buchi[\{w,s\}]$; the target states $w,s$ are indicated using double borders.
No Markov or finite-memory strategy for Max is $\eps$-optimal from $d$, for any $\eps<1$.
}
\label{fig:bad-match}
\end{minipage}%
\quad
\begin{minipage}{.47\textwidth}
  \centering
  \includegraphics{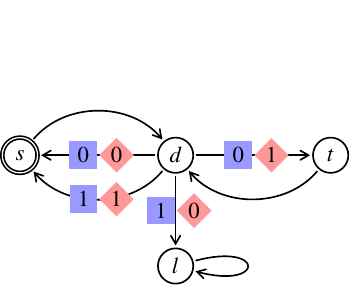}
\captionof{figure}{The Simplified Bad Match.
  The state $w$ is removed and the action pair $(1,1)$ leads to $s$ instead.
  The objective is $\Buchi[\{s\}]$, i.e., the target set $\{s\}$ is a singleton.
No Markov or finite-memory strategy for Max is $\eps$-optimal from $d$, for any $\eps<1$.
}
\label{fig:bad-match-simplified}
\end{minipage}
\end{figure}

The game $\game$ depicted in \cref{fig:bad-match}
is called the ``Bad Match''.
It was originally discussed with an expected $\limsup$ objective
with state-based rewards in $\{-1,0,1\}$
\cite{thuijsman1992optimality,MaitraSudderth:DiscreteGambling}.
Here we slightly adapt its presentation as a concurrent B\"uchi game.
There are action sets $A=B=\{0,1\}$ and the objective is $\Bu$ with target states~$\{s,w\}$.
  The initial state is $d$.
  The transition distributions are 
Dirac and satisfy
\(
\probp(d,1,1)(w)
= \probp(d,1,0)(l)
= \probp(d,0,1)(t)
= \probp(d,0,0)(s) = 1
 \);
the motion from states $w,l,s,t$ follows the unique depicted edge under all actions.
Intuitively, $w$ and $l$ are winning and losing sinks respectively, whereas
$s$ and $t$ represent a temporary win and loss
for Max, respectively.
The Simplified Bad Match in \Cref{fig:bad-match-simplified} is very similar,
but changes the permanent win $w$ to a temporary win in $s$,
so that the target set $\{s\}$ is a singleton.

In the Bad Match, all states except $l$ have value $1$ by
\cite[Lemma 13.8]{MaitraSudderth:DiscreteGambling};
see also \Cref{prop:bad-match-eps-opt-strat} below.
In the following
we consider the strategy complexity of the Bad Match
from Max's point of view.
First, Max does not have any optimal strategy from $d$, but only $\eps$-optimal strategies.

\begin{proposition}\label{prop:bad-match-no-opt}
In the Bad Match $\game$, Max does not have any optimal strategy from state $d$.
\end{proposition}
\begin{proof}{Proof.}
Either Max always plays action $0$, in which case Min can win by always
playing action $1$.
Or, otherwise, there exists some feasible finite history $h$ in which Max always plays
action $0$, after which Max plays action $1$ with some positive probability
$\delta>0$ for the first time.
Then Min can play such that $p \eqdef \probm_{\game,d,\zstrat,\ostrat}(hZ^\omega)>0$,
and then Min plays action $0$ after $h$.
This makes Max's attainment $\le 1-p\delta < 1$.
\hfill\Halmos
\end{proof}

The following two propositions state lower bounds for the complexity of
Max strategies:
neither finite-memory strategies nor Markov strategies for Max are $\eps$-optimal.

\begin{proposition}\label{claim:badmatch-finite}
In the Bad Match $\game$, 
every finite-memory Max strategy $\zstrat$
is worthless, i.e.,
$\inf_{\ostrat \in \ostratset}\probm_{\game,d,\zstrat,\ostrat}({\Bu})=0$.
\end{proposition}
\begin{proof}{Proof.}
Let $\zstrat$ be a Max strategy based on some finite memory set $M$, and fix an 
$\eps > 0$.
We show that Min can win with probability $1-\eps$,
i.e., that there is a Min strategy $\pi$ so that
$\probm_{\game,d,\zstrat,\ostrat}({\Bu})\le \eps$.

Choose $\eps_1, \eps_2 > 0$ with $\eps = \eps_1 + \eps_2$
and first consider the Min strategy, $\pi_1$, which in each step plays action $1$ with probability $\eps_1$ and action $0$ with probability $1-\eps_1$.
Then $\zstrat$ and $\pi_1$ induce a Markov chain with finite state space $S \x M$.
Since $\pi_1$ plays both actions with constant nonzero probabilities,
all states $(x,m)$ in a bottom strongly-connected component (BSCC) of this
Markov chain must satisfy one of the following properties.
Either $x \in \{w,l\}$ or $m$ is a Max memory mode from which Max can never
play action $1$ in the future.
By standard properties of finite-state Markov chains, there is a number $K$
s.t., independently of the start state, the probability of having 
entered a BSCC in the first $K$ steps is
$\ge 1-\eps_2$.
In particular, $K$ depends only on the induced Markov chain,
and thus only on $\game$ and the strategies $\zstrat$ and $\pi_1$.
Now we define Min's counter-strategy $\pi$, that depends on $K$ and
thus on $\zstrat$.
The Min strategy $\pi$ plays like $\pi_1$ in the first $K$ 
steps and action $1$ always thereafter.

Consider the strategies $\zstrat$ and $\ostrat$.
The probability that within $K$ steps the permanently winning state $w$ is 
reached is $\le \eps_1$; this follows from the definition of $\pi_1$.
The probability that within $K$ steps the finite-state Markov chain has not yet 
reached a BSCC is $\le \eps_2$.
It follows that the probability that within $K$ steps Max has permanently 
lost or reaches a memory mode from where he will never 
play action $1$ is at least $1-\eps_1-\eps_2 = 1-\eps$.
Thus, Min wins with probability at least $1-\eps$.
\hfill\Halmos
\end{proof}

Notice that \Cref{claim:badmatch-finite}
even holds for all private finite-memory Max strategies.

It was shown in
\cite[Lemma 13.9]{MaitraSudderth:DiscreteGambling}
that also all Markov strategies are worthless for Max.

\begin{restatable}[{\cite[Lemma 13.9]{MaitraSudderth:DiscreteGambling}}]{proposition}{propbadmatchnomarkov}\label{prop:bad-match-no-markov}
In the Bad Match $\game$, 
every Markov strategy $\zstrat$ for Max
is worthless, i.e.,
$\inf_{\ostrat \in \ostratset}\probm_{\game,d,\zstrat,\ostrat}({\Bu})=0$.
\end{restatable}
\begin{proof}{Proof.}
Let $\sigma$ be a Markov strategy for Max and for any step $n\ge0$,
let $r_n$ be the probability with which $\sigma$ chooses action $1$
at step $2n$ (in state $d$).
There are two cases, whether or not the sum of $r_n$
converges or diverges as $n$ goes to $\infty$.

Case $\sum_{n=0}^{\infty}r_n=\infty$.
Then almost surely, there are infinitely many steps in which Max chooses action $1$.
If Min's strategy $\ostrat$
always chooses action $0$, she guarantees that the game almost surely reaches the losing sink $l$. Thus,
$\probm_{\game,d,\zstrat,\ostrat}({\Bu})=0$.

Case $\sum_{n=0}^{\infty}r_n<\infty$.
Let $\eps>0$. 
Then we can choose $K$ so that $\sum_{n=K}^{\infty}r_n\le \eps$.
In particular, $\eps$ bounds the probability with which Max ever plays action $1$ after step $K$.
Let $\pi$ be the Min strategy that chooses $0$ until step $K$ and $1$ thereafter.
This guarantees that until step $K$, the game will surely avoid the winning sink $w$.
If in the first $K$ steps, Max plays action $1$ then the game moves to the losing sink $l$ and is won by Min.
Otherwise, if the game does not reach $l$ within the first $K$ steps,
then the probability that Max always plays $0$, and the play therefore alternates between non-target states $d$ and $t$ indefinitely, is at least $1-\eps$.
Hence, $\probm_{\game,d,\zstrat,\ostrat}({\Bu})\le \eps$.
\hfill\Halmos
\end{proof}

\begin{remark}\label{rem:bad-match-simplified}
In the Simplified Bad Match of \Cref{fig:bad-match-simplified},
state $d$ still has value $1$ (\Cref{prop:bad-match-eps-opt-strat}),
but all finite-memory strategies
and Markov strategies for Max are worthless.
Since the modification relative to the Bad Match only makes
it harder for Max,
\Cref{prop:bad-match-no-opt,claim:badmatch-finite,prop:bad-match-no-markov}
carry over.
\end{remark}  

We now show that there exists an $\eps$-optimal Max strategy
in the (Simplified) Bad Match that uses just a step counter plus one extra bit of public memory.

\begin{proposition}\label{prop:bad-match-eps-opt-strat}
The value of state $d$ in the (Simplified) Bad Match is $1$
and for every $\eps>0$ Max has an $\eps$-optimal 1-bit Markov strategy.
\end{proposition}
\begin{proof}{Proof.}
For the Bad Match,
\citeauthor{MaitraSudderth:DiscreteGambling}
\cite[Lemma 13.8]{MaitraSudderth:DiscreteGambling} prove
that state $d$ has value $1$ by showing that no Min strategy can enforce
a value strictly below $1$ (i.e., the upper value of $d$ is $1$)
and then appealing to the determinacy of the game.

We now give an alternative proof, even for the Simplified Bad Match,
by explicitly constructing 
a Max strategy $\zstrat$ that attains $\ge 1-\eps$ from $d$.
Afterwards we argue how $\zstrat$ can be modified to give a 1-bit Markov strategy.

First, recall that finite-state concurrent games with \emph{reachability} objectives
always admit memoryless randomized $\eps$-optimal strategies
\cite[Section 7.7]{MaitraSudderth:DiscreteGambling}.
In particular, in the Simplified Bad Match of \cref{fig:bad-match-simplified},
Max has such a strategy to reach the target set $\{s\}$ from $d$ with
probability $\ge 1-\eps$.
He can do so by playing the following mixed action every time
the play reaches state $d$:
play action $1$ with probability $\eps$ and
action $0$ with probability $1-\eps$. 
Indeed, consider this Max strategy and an arbitrary Min strategy.
Let $E$ be the event that Min eventually plays action $0$.
Conditioned under $E$, there is a first time that Min plays action $0$.
Then either Max has already played action $1$ before, in which case he has
already won by reaching state $s$, or not,
in which case Max plays action $0$ with probability $1-\eps$,
so he wins via state $s$ with probability $1-\eps$.
Therefore, conditioned under $E$, Max wins with probability $\ge 1-\eps$.
On the other hand, conditioned under not-$E$, Min always plays action $1$
and thus Max wins almost surely by eventually reaching state $s$.
Thus Max wins by reaching state $s$ with probability at least $1-\eps$.

Now consider the following Max strategy $\sigma$ from $d$ for the B\"uchi
objective.
It proceeds in phases $1,2,3,\dots$, separated by visits to state $s$
(unless the losing state $l$ has been reached).
In every phase $i$, Max plays an $\eps_i$-optimal strategy towards reaching
the set $\{s\}$. 
When the play returns from $s$ to $d$ then the next phase $i+1$ begins.
By choosing $\eps_i \eqdef \eps 2^{-i}$, we obtain that the B\"uchi
objective is satisfied with probability $\ge 1 - \sum_i \eps_i = 1-\eps$.
This strategy $\sigma$ uses infinite memory, since it needs to keep track of
the current phase $i$ (and this cannot be done with just a step counter by
\Cref{prop:bad-match-no-markov}).

We construct a Max strategy $\sigma'$ that is similar to $\sigma$.
It also works in phases, but each phase $i$ has a pre-determined length $l_i$.
In each phase $i$, Max plays a memoryless $\eps_i$-optimal strategy towards reaching
the target set $\{s\}$, and the length $l_i$ is chosen sufficiently large such that
the probability of reaching the target before the end of the phase is
$\ge 1-\eps_i-\eps_i'$ for some sufficiently small $\eps_i'>0$.
After reaching $s$, for the rest of the current phase, Max plays conservatively by
always choosing action $0$. Then the next phase $i+1$ begins, etc.
By choosing $\eps_i \eqdef \eps_i' \eqdef \eps 2^{-(i+1)}$, we obtain that the B\"uchi
objective is satisfied with probability $\ge 1-\eps$.
The strategy $\sigma'$ is a 1-bit Markov strategy. It uses the step counter to
determine the current phase, since the lengths $l_i$ are pre-determined,
and it uses 1 bit of public memory to remember whether it has already visited state $s$ in
the current phase and thus needs to play conservatively until the end of this
phase. The 1 bit ensures that $\sigma'$ loses
$\le \eps_i + \eps_i' = \eps 2^{-i}$ only \emph{once} in each phase $i$, and not
multiple times per phase.

The 1-bit Markov strategy $\sigma'$ also works in the original Bad
Match (with target set $\{w,s\}$),
since the difference between the games only benefits Max (permanent win in
$w$ on action pair
$(1,1)$ instead of just a temporary win in $s$).
\hfill\Halmos
\end{proof}

We show, as \cref{thm:Buchi}, that such an $\eps$-optimal 1-bit
Markov strategy for Max \emph{always} exists, even in countably infinite-state
stochastic B\"uchi games; see \Cref{sec-buchi}.

\mysec{Turn-based Stochastic B\"uchi Games}\label{subsec:turn-based-buchi}

The counterexamples from \Cref{fig:bad-match,fig:bad-match-simplified}
are finite-state concurrent B\"uchi games.
The lower bounds on Max's strategy complexity
(\Cref{claim:badmatch-finite,prop:bad-match-no-markov})
do \emph{not} carry over to
finite-state \emph{turn-based} B\"uchi games (where Max always has optimal memoryless
deterministic strategies for reachability \cite{CONDON1992203S} and thus also
for B\"uchi objectives). 
However, these lower bounds do carry over to turn-based B\"uchi games
with a \emph{countably infinite} number of states, even if the target set is
just a singleton.
This can be shown by constructing an infinite-state turn-based version of the
Simplified Bad Match with a singleton target set.

\begin{definition}[Turn-based Simplified Bad Match]\label{def:turn-based-bad-match}
We define an infinite-state turn-based version $\hat{\game}$
of the Simplified Bad Match from \Cref{fig:bad-match-simplified}.
The state $d$ is split into
an infinite chain of states $d_0, d_1, d_2, \dots$ where $d=d_0$.
At every state $d_i$, Max has two available actions, and only Max's action
affects the outcome. For the first action the game goes to $d_{i+1}$,
and for the second action the game goes to $e_i$.
Intuitively, by going to state $e_i$, Max chooses a mixed action
in the original Simplified Bad Match
that plays the action $1$
with probability $2^{-i}$ and the action $0$ with probability $1-2^{-i}$.

At every state $e_i$, Min's available actions are $\{0,1\}$
and only Min's action matters.
If Min chooses action $0$ then the game goes to state $(1,0)$
with probability $2^{-i}$
and to state $(0,0)$ with probability $1-2^{-i}$.
If Min chooses action $1$ then the game goes to state $(1,1)$ with probability $2^{-i}$
and to state $(0,1)$ with probability $1-2^{-i}$.
This mimics the chosen actions in the original Simplified Bad Match.

Moreover, there are transitions $(1,1) \to s$, $(0,0) \to s$,
$(0,1) \to t$ and $(1,0) \to l$, encoding the outcomes of these action pairs
in the Simplified Bad Match.
As before, states $s$ and $t$ have transitions back to $d$, while $l$ is a
losing sink state.
The only target state is $s$.
\end{definition}

Lifting \Cref{prop:bad-match-no-markov} to the turn-based Simplified Bad Match $\hat{\game}$ is
not immediate. Traversing arbitrarily long prefixes of
the chain of states $d_0, d_1, d_2, \dots$
in $\hat{\game}$
takes time, and thus one cannot simply carry Markov strategies from
$\hat{\game}$ to $\game$ (unlike for finite-memory strategies).
Still, Markov strategies are worthless for Max in $\hat{\game}$.

\begin{restatable}{proposition}{propbadmatchnomarkovturnbased}\label{prop:bad-match-no-markov-turnbased}
  In the
  turn-based Simplified Bad Match $\hat{\game}$
  of Def.~\ref{def:turn-based-bad-match},
  state $d$ has value $1$, but Max does not have any optimal
  strategy from $d$.
  
  Moreover, every finite-memory strategy and every Markov strategy $\zstrat$ for Max
  is worthless, i.e.,
$\inf_{\ostrat \in \ostratset}\probm_{\hat{\game},d,\zstrat,\ostrat}({\Bu})=0$.
\end{restatable}
\begin{proof}{Proof.}
Recall that in countable concurrent games with finite action sets,
Max has memoryless $\eps$-optimal strategies for \emph{reachability}
objectives \cite[Corollary 3.9]{Secchi97}.
In particular, in the turn-based Simplified Bad Match of \Cref{def:turn-based-bad-match},
Max has such a strategy to reach the target set $\{s\}$ from $d$ with
probability $\ge 1-2^{-i}$ for any $i \in \N$,
by always going to state $e_i$.
Now consider the following Max strategy $\zstrat$ from $d$ for the B\"uchi
objective.
It proceeds in phases $1,2,3,\dots$, separated by visits to state $s$
(unless the losing state $l$ has been reached).
In every phase $i$, Max plays an $\eps_i$-optimal strategy towards reaching
the set $\{s\}$. 
When the play returns from $s$ to $d$ then the next phase $i+1$ begins.
For any $j \in \N$,
by choosing $\eps_i \eqdef 2^{-j} 2^{-i}$, we obtain that the B\"uchi
objective is satisfied with probability $\ge 1-2^{-j}$.
Since $j$ can be chosen arbitrarily large, it follows that $d$ has value $1$.

However, Max does not have any optimal strategy from state $d$.
Plays that always stay in the states $\{d_i\}_{i \in \N}$ attain nothing.
On the other hand, if Max ever enters some state $e_i$ then Min can limit
Max's attainment to $\le 1-2^{-i}$ by playing action $0$.
  
Now we consider finite-memory Max strategies.
  Towards a contradiction assume that there exists an
  $\eps$-optimal finite-memory Max strategy $\zstrat$ in $\hat{\game}$.
This would induce an $\eps$-optimal finite-memory Max strategy $\zstrat'$ in the Bad
Match $\game$ as follows.
Consider $\zstrat$ in some memory mode $m$ at state $d$, and let $p_i$ be the
probability that the next state in the set $\{e_i \mid i\ge 0\}$ that is
visited is $e_i$. Without restriction we assume that $\sum_{i \ge 0} p_i = 1$, since
staying in the states $\{d_i \mid i \ge 0\}$ forever is losing.
Now let $p(m) \eqdef \sum_i p_i2^{-i}$, and $\zstrat'$ in memory mode $m$ at
state $d$ plays action $1$ with probability $p(m)$.
Likewise, $\zstrat'$ updates its memory based on the observed pair of actions
in the same way as $\zstrat$ updates its memory based on the observed states
$(0,0), (0,1), (1,0), (1,1)$.
Then $\zstrat'$ is $\eps$-optimal in $\game$, which contradicts
\Cref{claim:badmatch-finite}.
(There exist other counterexamples. Even in countably infinite MDPs with a singleton target set,
finite-memory strategies are worthless for B\"uchi objectives, e.g.,
\cite[Fig.~2b]{KMSW2017}.)

Now we consider Markov strategies.  
To simplify the presentation, we consider a slightly modified game $\hat{\game}'$
where state $l$ is not a sink but instead has a transition back to $d$.
To compensate, we consider a modified objective $\Bu'$ which is defined as
``$\Bu$ and never visit state $l$''.
(I.e., plays that visit $l$ are still losing for Max.)
Thus it suffices to show that
$\inf_{\ostrat \in \ostratset}\probm_{\hat{\game}',d,\zstrat,\ostrat}({\Bu'})=0$
for every Markov strategy $\zstrat$ for Max from state $d$.

Let $\zstrat$ be a Max Markov strategy from state $d$
and let
$\eps >0$.
We will construct a Min strategy $\hat{\ostrat}$ such that 
$\probm_{\hat{\game}',d,\zstrat,\hat{\ostrat}}({\Bu'}) \le 2\eps$.
Since $\eps$ can be chosen arbitrarily small, the result follows.

We define $D \eqdef \{d_i \mid i \ge 0\}$ and
$E \eqdef \{e_i \mid i \ge 0\}$.
Without restriction we assume that no plays compatible with $\zstrat$
eventually stay in the set
of states $D$ forever, since those plays are losing and Max always has the
option to leave $D$.
Thus all plays compatible with $\zstrat$ (and any Min strategy $\ostrat$)
visit the set $E$, and thus state $d$, infinitely often.

Let $\ostrat_0$ (resp.\ $\ostrat_1$)
be the MD Min strategy that at every state $e_i$ chooses action $0$
(resp.\ action $1$).
By fixing the Max strategy $\zstrat$ and the Min strategy $\ostrat_0$ (resp.\ $\ostrat_1$)
in $\hat{\game}'$ we obtain the Markov chain $A_0$ (resp.\ $A_1$).
The states in $A_0$ (and $A_1$) have the form $(x,t)$ where $x$ is a state in $\hat{\game}'$
and $t \in \N$ is the time.
Thus $A_0$ and $A_1$ are countably infinite acyclic Markov chains.

The Markov strategy $\zstrat$ cannot remember Min's past actions, since
all paths from $e_i$ back to $d$ have the same length $3$,
and thus Max's decisions do not depend on whether Min plays $\ostrat_0$
or $\ostrat_1$. Thus the Markov chains $A_0$ and $A_1$ are very similar.

Let $N \subseteq \N$ be the set of times $t$ such that state $(d,t)$ is
visited with nonzero probability in $A_0$.
Consider the event
$
R \eqdef \{\rho \mid \mbox{$\rho$ visits the set
  $\{(1,0),(1,1)\}$ infinitely often}\}
$.
Intuitively, the event $R$ consists of the plays where Max
has played the action $1$ infinitely often.

Since $R$ is shift-invariant, it follows from L\'{e}vy's 0-1 law that
the probability of satisfying $R$ almost surely converges to either $0$
or $1$. 
Since runs in $A_0$ visit states of the form $(d,t)$ infinitely often (at 
strictly increasing times $t \in N$), it follows that
there is a threshold $n_\eps \in \N$ such that
for each $t \in N$ with $t \ge n_\eps$ either
$\probm_{A_0,(d,t)}(R) \le \eps$ or $\probm_{A_0,(d,t)}(R) \ge 1-\eps$.
These probabilities need not be monotone in $t$.
However, the satisfaction of $R$ does not depend on whether $\ostrat=\ostrat_0$ or
$\ostrat=\ostrat_1$, and thus these inequalities (for the same $t$)
also hold if $A_0$ is exchanged with $A_1$.
Let $T_1$ (resp.\ $T_2$) denote the set of times
$t \in N$ with $t \ge n_\eps$ where
$\probm_{A_0,(d,t)}(R) \le \eps$ and
$\probm_{A_1,(d,t)}(R) \le \eps$ 
(resp.\ $\probm_{A_0,(d,t)}(R) \ge 1-\eps$
and $\probm_{A_1,(d,t)}(R) \ge 1-\eps$).

Moreover, conditioned under $\overline{R}$, the probability of ever visiting
$\{(1,0),(1,1)\}$ after time $t$ must converge to zero as $t \to \infty$.
Therefore
$\lim_{N \ni t\to \infty}\probm_{A_0,(d,t)}(\reach[\{(1,0),(1,1)\}] \cap \overline{R})=0$
and
$\lim_{N \ni t\to \infty}\probm_{A_1,(d,t)}(\reach[\{(1,0),(1,1)\}] \cap
\overline{R})=0$.
Thus there exists a number $m_\eps \ge n_\eps$ such that for all
$N \ni t \ge m_\eps$ we have both
$\probm_{A_0,(d,t)}(\reach[\{(1,0),(1,1)\}] \cap \overline{R}) \le \eps$
and
$\probm_{A_1,(d,t)}(\reach[\{(1,0),(1,1)\}] \cap \overline{R}) \le \eps$.

We now define the Min strategy $\hat{\ostrat}$.
It plays $\ostrat_0$
until the game reaches a state $(d,t)$ with $t \ge m_\eps$
for the first time (which implies that $t \in N$).
If $t \in T_1$ then $\hat{\ostrat}$ plays $\ostrat_1$
henceforth.
Otherwise, if $t \in T_2$, then $\hat{\ostrat}$ continues playing $\ostrat_0$
forever.

In the former case, $t \in T_1$, we have
$\probm_{\hat{\game}',d,\zstrat[t],\hat{\ostrat}}({\Bu'})
\le
\probm_{A_1,(d,t)}(R)
+
\probm_{A_1,(d,t)}(\reach[\{(1,0),(1,1)\}] \cap \overline{R})
\le
\eps
+
\eps
= 2\eps$.

In the latter case, $t \in T_2$, we have
$\probm_{\hat{\game}',d,\zstrat[t],\hat{\ostrat}}({\Bu'})
\le
\probm_{A_0,(d,t)}(\overline{R})
+
\probm_{A_0,(d,t)}(\Bu' \mid R)
\le
\eps
+
\probm_{A_0,(d,t)}(\neg\reach[\{l\}] \mid R))
=\eps$.
Thus
$\probm_{\hat{\game}',d,\zstrat,\hat{\ostrat}}({\Bu'}) \le
\max(2\eps,\eps) = 2\eps$ as required.
\hfill\Halmos
\end{proof}

\Cref{prop:bad-match-no-markov-turnbased}
shows that finite-memory strategies and Markov strategies
are worthless for Max in turn-based countable B\"uchi games
with a \emph{singleton target set}.
This is orthogonal to a
previous result in \cite{KMST:ICALP2019} that Markov strategies
are worthless for B\"uchi objectives in countably infinite
MDPs, because the counterexample in \cite{KMST:ICALP2019}
uses an \emph{infinite} target set.
The strategy complexity of B\"uchi objectives in countably infinite
MDPs with a \emph{singleton target set} (or a finite target set) is open.

\newpage
\section{Transience Games}
\label{sec-trans}
\begin{figure*}[t]
  \begin{center}
   \includegraphics{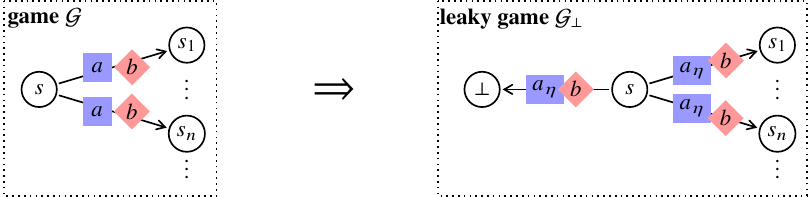} 
\end{center}	
\caption{In the game~$\game$, the  distribution~$p(s,a,b)$ is such that $\sum_{i\in \nat} p(s,a,b)(s_i)=1$. 
In the leaky game~$\game_{\bot}$, for every rational number~$\eta\in (0,1)$, the probability distribution~$p_{\bot}(s,a_{\eta},b)$ is defined such that the support $(s,a,b)$ also includes~$\bot$. More precisely, we have $p_{\bot}(s,a_{\eta},b)(s_i) = (1-\eta)p(s,a,b)(s_i)$ and $p_{\bot}(s,a_{\eta},b)(\bot)=\eta$. 
}
\label{fig:leakygame}
\end{figure*}

In this section we prove that Max has memoryless strategies that are uniformly
$\eps$-optimal for Transience.

\thmTr*

We now fix a game~$\game$  with the set~$S$ of states and with the
transience objective $\Tr$. From~$\game$,
we construct a \emph{leaky} game~$\game_\bot$, described below,
where $\bot\not \in S$ is a fresh losing sink state.
Denote by $\avoid[\bot]$ the event of not visiting~$\bot$.
The goal of the construction is that, for every finite-memory strategy of Max
and any (general) strategy of Min, the events $\Tr$ and
$\avoid[\bot]$ are equal up to measure zero.
The intuition is that it then suffices to construct strategies
in $\game_\bot$ that focus on avoiding $\bot$, which is conceptually easier.

Towards this goal,
for every state~$s \in S$
and every Max action~$a \in A(s)$,
we replace this Max action $a$
by a countably infinite collection of Max actions
$a_\eta$ for every \emph{rational} number~$\eta\in (0,1)$.
The Max 
action $a_\eta$ leads with probability~$\eta$ (regardless of Min's action)
to~$\bot$, and with probability $1{-}\eta$ behaves like the original Max
action $a$; see~\Cref{fig:leakygame} for an illustration of the construction.
Hence the set of Max actions in~$\game_{\bot}$ is countably infinite. 
Below, we show that for all events $E\subseteq S^{\omega}$
(note that $\bot \not \in S$) and states $s$,
the values of~$s$ for $E$ in games~$\game$ and~$\game_{\bot}$ are equal.
In particular, the equality  holds for the transience objective~$\Tr$.

Let us first show one of the required inequalities.

\begin{lemma} \label{lem:leaky-one-direction}
For all measurable events~$E\subseteq S^{\omega}$ and all states~$s\in\states$,
$
  \valueof{\game,E}{s} \le \valueof{\game_{\bot},E}{s}\,.
$
\end{lemma}
\begin{proof}{Proof.}
Given a rational $\eps>0$ and any Max strategy~$\sigma$ in~$\game$ we derive a Max strategy~$\sigma_{\bot}$ in $\game_{\bot}$, with only an additional $\eps$ error for $E$. 
Let $\sigma_{\bot}$ be such that 
$\sigma_{\bot}(h_{\bot})(a_{\eta}) = \sigma(h)(a)$ where  $\eta=\eps\cdot 2^{-(|h_{\bot}|+1)}$, where 
$h$ is the history in~$\game$ corresponding to~$h_{\bot}$ in $G_{\bot}$ (i.e., every Max action~$a_q$, with $q\in (0,1)$, in $h_{\bot}$ is replaced with the action~$a$ in $h$).
Then, by a union bound, we have for all Min strategies~$\pi_\bot$,
\[
\probm_{\game_\bot,s,\sigma_{\bot}, \pi_\bot}(\avoid[\bot]) 
\, \ge \, 1-\sum_{i=1}^\infty \frac{\eps}{2^i}
\, \ge \, 1-\eps\,.
\]

Let $\pi_\bot$ be an arbitrary Min strategy in~$\game_\bot$.
Define $\pi$, a Min strategy in~$\game$, so that $\pi$~acts on any history~$h$ in~$\game$ in the way that $\pi_\bot$ acts on the corresponding history~$h_\bot$ in~$\game_\bot$ assuming that Max plays~$\sigma_\bot$ (i.e., every Max action $a$ in~$h$ is replaced with the action $a_\eta$ for the $\eta$ defined above for~$\sigma_\bot$).
Since $E \subseteq \avoid[\bot]$, 
\begin{align*}
\probm_{\game_\bot,s,\sigma_{\bot}, \pi_\bot}(E)\,
&= \, \probm_{\game_\bot,s,\sigma_{\bot}, \pi_\bot}(E \mid \avoid[\bot]) \,  \probm_{\game_\bot,s,\sigma_{\bot}, \pi_\bot}(\avoid[\bot])\\
&= \, \probm_{\game,s,\sigma, \pi}(E) \, \probm_{\game_\bot,s,\sigma_{\bot}, \pi_\bot}(\avoid[\bot]) \\
&\geq \, \probm_{\game,s,\sigma, \pi}(E) (1-\eps) \,.
\end{align*}
For all $i\geq 1$, every (multiplicatively) $2^{-i}$-optimal strategy~$\sigma$ in $\game$ 
transferred to $\game_{\bot}$ in the described way above yields a 
(multiplicatively) $2^{-i+1}$-optimal strategy~$\sigma_{\bot}$ in $\game_{\bot}$, concluding the proof. 
\hfill\Halmos
\end{proof}

Now we show that the values are in fact equal and that, moreover, Max strategies in $\game_{\bot}$ can be carried back to $\game$.

\begin{lemma}\label{col:carrybacktoG}
Let $E\subseteq S^{\omega}$ be a measurable event. For all states~$s\in\states$ we have
$
  \valueof{\game,E}{s}=\valueof{\game_{\bot},E}{s}\,.
$
Moreover, let $\eps > 0$.
Let $\sigma_{\bot}$ be an $M$-based Max strategy in $\game_{\bot}$, that is multiplicatively $\eps$-optimal from 
every state (with a fixed initial mode).  
There exists an $M$-based  Max strategy~$\sigma$ in $\game$, that is multiplicatively $\eps$-optimal from 
every state (with a fixed initial mode).     
\end{lemma}
\begin{proof}{Proof.}
Let $\game'$ denote the following ``intermediate'' (between $\game$ and $\game_\bot$) game: $\game'$~is like $\game_\bot$ except that each action $a_\eta$ is equivalent to action~$a$ in~$\game$ and thus non-leaky; i.e., $p'(s,a_\eta,b) = p(s,a,b)$.

Let $E\subseteq S^{\omega}$ be an event, and let $\eps > 0$.
Let $\sigma_{\bot}$ be an $M$-based Max strategy in $\game_{\bot}$, with initial memory mode~$m_0$.
We assume that $\sigma_\bot$ is multiplicatively $\eps$-optimal in~$\game_\bot$ from every state; i.e., for every initial state~$s_0$ and for every Min strategy~$\pi$ we have $\probm_{\game_\bot,s_0,\sigma_{\bot}, \pi}(E) \ge \valueof{\game_{\bot},E}{s_0} \cdot (1 - \eps)$.
Since the state~$\bot \not\in \states$ is losing, the \emph{same} Max strategy~$\sigma_\bot$, which we also name $\sigma' := \sigma_\bot$ for notational clarity later on, performs at least as well in~$\game'$; i.e., we have
\begin{equation} \label{eq:carrybacktoG}
\probm_{\game',s_0,\sigma', \pi'}(E) \ \ge \ \valueof{\game_\bot,E}{s_0} \cdot (1 - \eps) \qquad \text{for all $s_0$ and all~$\pi'$.} 
\end{equation}

Next we describe how we carry the strategy $\sigma'$ in~$\game'$ back to~$\game$, without changing the memory~$M$.
Write $A':=\{a_{\eta}: a\in A, \eta\in \Q\cap (0,1)\}$.
Suppose $\sigma'$ has the action and update functions $\stratAct'  : \states \x M \to \dist(A')$ and $\stratUp' : \states\x M \x A'\x B\x \states \to \dist(M)$, respectively.
We define the strategy~$\sigma$ via its action and update functions $\stratAct$ and~$\stratUp$ as follows:
\begin{align*}
\stratAct(s,m)(a) \ &\eqdef \ \sum_{\eta \in \Q \cap (0,1)} \stratAct'(s,m)(a_\eta) \;, \\
\stratUp(s,m,a,b,s')(m') \ &\eqdef \ \frac{\sum_{\eta \in \Q \cap (0,1)} \stratAct'(s,m)(a_\eta) \cdot \stratUp'(s,m,a_\eta,b,s')(m')}{\sum_{\eta \in \Q \cap (0,1)} \stratAct'(s,m)(a_\eta)}\;.
\end{align*}

Note that
\begin{equation} \label{eq:carrybacktoG-product}
\stratAct(s,m)(a) \cdot \stratUp(s,m,a,b,s')(m') \ = \  \sum_{\eta \in \Q \cap (0,1)} \stratAct'(s,m)(a_\eta) \cdot \stratUp'(s,m,a_\eta,b,s')(m')\;.
\end{equation}
Let $s_0$ be an arbitrary initial state, and let $\pi$ be an arbitrary Min strategy in~$\game$.
Define a Min strategy~$\pi'$ in~$\game'$ that works like~$\pi$, treating any Max action $a_\eta$ from~$\game'$ as $\pi$~treats Max action~$a$ from~$\game$.
We show that for any ``expanded'' (i.e., taking Max's memory into account) history~$h$ that starts with $s_0 m_0$, i.e., for every $n \ge 0$ and every $s_0 m_0 a_1 b_1 s_1 m_1 \cdots a_n b_n s_n m_n \in \{(s_0,m_0)\}\times (A \times B \times S \times M)^n$, we have
\begin{equation} \label{eq:carrybacktoG-ind}
\begin{aligned}
     &\probm_{\game,s_0,\sigma,\pi}(s_0 m_0 a_1 b_1 s_1 m_1 \cdots a_n b_n s_n m_n)  \\
= \  &\sum_{\eta_1, \ldots, \eta_n \in \Q \cap (0,1)} \probm_{\game',s_0,\sigma',\pi'}(s_0 m_0 (a_1)_{\eta_1} b_1 s_1 m_1 \cdots (a_n)_{\eta_n} b_n s_n m_n)\;,
\end{aligned}
\end{equation}
where by the probability of a finite history we mean the probability of its corresponding cylinder set.
We prove~\cref{eq:carrybacktoG-ind} by induction on~$n$.
Concerning the induction base, for $n=0$ both sides are equal to~$1$.
Concerning the step, suppose \eqref{eq:carrybacktoG-ind} holds for some~$n \ge 0$.
We have
\begin{align*}
    & \probm_{\game,s_0,\sigma,\pi}(s_0 m_0 a_1 b_1 s_1 m_1 \cdots a_n b_n s_n m_n a_{n+1} b_{n+1} s_{n+1} m_{n+1})  \\
= \ & \probm_{\game,s_0,\sigma,\pi}(s_0 m_0 a_1 b_1 s_1 m_1 \cdots a_n b_n s_n m_n) \cdot \stratAct(s_n,m_n)(a_{n+1}) \cdot \mbox{} \\ 
    & \mbox{} \cdot \pi(s_0 m_0 a_1 b_1 s_1 m_1 \cdots a_n b_n s_n m_n)(b_{n+1}) \cdot p(s_n,a_{n+1},b_{n+1})(s_{n+1}) \cdot \mbox{} \\
    & \mbox{} \cdot \stratUp(s_n,m_n,a_{n+1},b_{n+1},s_{n+1})(m_{n+1})\;.
\end{align*}
To continue this equality chain, we will use the induction hypothesis~\cref{eq:carrybacktoG-ind} and~\cref{eq:carrybacktoG-product}.
We also use the definition of~$\pi'$ (treat each~$a_\eta$ as $\pi$~treats~$a$) and the definition of~$\game'$ (which implies $p(s,a,b) = p'(s,a_\eta,b)$ for each~$\eta$).
Thus, we obtain
\begin{align*}
    & \probm_{\game,s_0,\sigma,\pi}(s_0 m_0 a_1 b_1 s_1 m_1 \cdots a_n b_n s_n m_n a_{n+1} b_{n+1} s_{n+1} m_{n+1})  \\
=  & \sum_{\eta_1, \ldots, \eta_n \in \Q \cap (0,1)} \probm_{\game',s_0,\sigma',\pi'}(s_0 m_0 (a_1)_{\eta_1} b_1 s_1 m_1 \cdots (a_n)_{\eta_n} b_n s_n m_n) \cdot \mbox{} \\
    & \hspace{15mm} \cdot \hspace{-5mm}\sum_{\eta_{n+1} \in \Q \cap (0,1)} \hspace{-2mm}\stratAct'(s_n,m_n)((a_{n+1})_{\eta_{n+1}}) \cdot \stratUp'(s_n,m_n,(a_{n+1})_{\eta_{n+1}},b_{n+1},s_{n+1})(m_{n+1}) \cdot \mbox{} \\
    & \hspace{18mm} \qquad \mbox{} \cdot \pi'(s_0 m_0 (a_1)_{\eta_1} b_1 s_1 m_1 \cdots (a_n)_{\eta_n} b_n s_n m_n)(b_{n+1}) \cdot p'(s_n,(a_{n+1})_{\eta_{n+1}},b_{n+1})(s_{n+1}) \\
=  &\sum_{\eta_1, \ldots, \eta_n, \eta_{n+1} \in \Q \cap (0,1)} \probm_{\game',s_0,\sigma',\pi'}(s_0 m_0 (a_1)_{\eta_1} b_1 s_1 m_1 \cdots (a_n)_{\eta_n} b_n s_n m_n (a_{n+1})_{\eta_{n+1}} b_{n+1} s_{n+1} m_{n+1})\;,
\end{align*}
which completes the induction step.
Thus, we have established \cref{eq:carrybacktoG-ind}.
Hence, the probability measures $\probm_{\game,s_0,\sigma,\pi}$ and $\probm_{\game',s_0,\sigma',\pi'}$ coincide when projected to state sequences; in particular, we have $\probm_{\game,s_0,\sigma,\pi}(E) = \probm_{\game',s_0,\sigma',\pi'}(E)$.
With \cref{eq:carrybacktoG} it follows:
\begin{equation}\label{eq:carrybacktoG-end}
\probm_{\game,s_0,\sigma, \pi}(E) \ \ge \ \valueof{\game_\bot,E}{s_0} \cdot (1 - \eps) \,.
\end{equation}
Since $s_0$ and~$\pi$ were chosen arbitrarily, it follows that $\valueof{\game,E}{s_0} \ge \valueof{\game_\bot,E}{s_0}$.
Hence, using \cref{lem:leaky-one-direction},
\[
\valueof{\game,E}{s_0} \ = \ \valueof{\game_\bot,E}{s_0} \,,
\]
as required.
By \cref{eq:carrybacktoG-end} we obtain
\[
\probm_{\game,s_0,\sigma, \pi}(E) \ \ge \ \valueof{\game,E}{s_0} \cdot (1 - \eps) \,,
\]
which proves that $\sigma$~is multiplicatively $\eps$-optimal, as required.

We remark that in this proof we have assumed that Max's memory modes are public, i.e., Min can take into account Max's memory mode.
By replacing terms like $\pi(s_0 m_0 a_1 b_1 s_1 m_1 \cdots a_n b_n s_n m_n)$ with $\pi(s_0 a_1 b_1 s_1 \cdots a_n b_n s_n)$, a very similar proof shows the analogous statement for private memory.
\hfill\Halmos 
\end{proof}

The leaky game $\game_{\bot}$ has the following important property
for finite-memory Max strategies (and thus in particular for
memoryless Max strategies).

\begin{lemma}\label{lem:strongerTrNotbot} 
In $\game_{\bot}$, for all finite-memory Max strategies~$\sigma$, all Min
strategies~$\pi$ and states~$s$,
the events $\Tr$ and $\avoid[\bot]$ differ only by a nullset. Formally,
\[
\probm_{\game_{\bot},s,\sigma, \pi} ((\Tr \setminus \avoid[\bot]) \cup 
(\avoid[\bot] \setminus \Tr))=0\,.
\]
\end{lemma}
\begin{proof}{Proof.}
Consider  the set~$\{1,2,\cdots\}$ of states in $\game_{\bot}$, excluding~$\bot$.   
Write $E^n(i)$ for the event that the state~$i$ is visited at least~$n$ times. 
The event~$E^{\infty}(i)$, defined by
\[
  E^{\infty}(i) \eqdef \bigcap_{n \in \N} E^n(i)\, ,
\]
is the event that state~$i$ is visited infinitely often.
Let  $\uplus$ denote the disjoint union.
Since $\bot$ is absorbing, 
\begin{align*}
	\avoid[\bot] = \, \Tr\,  \uplus \, \bigcup_{i \in \N} \, E^{\infty}(i) \,.
\end{align*}	

It suffices to show
that, for all~$i \in \N$, 
$
  \probm_{\game_{\bot},s,\sigma, \pi} (E^{\infty}(i) )=0.
$
Let $M$ be the \emph{finite} set of memory modes of the Max strategy $\sigma$.
By definition,
$E^{n+1}(i) \subseteq E^n(i)$ holds for all $n\in \nat$. Then it follows that 
\begin{align*}
	\probm_{\game_{\bot},s,\sigma, \pi} (E^{\infty}(i)) = &
	\lim_{n\to \infty }\probm_{\game_{\bot},s,\sigma, \pi} (E^{n}(i)) &
                                                                              \text{by
                                                                              continuity
  of measures}\\
	\leq & \lim_{n\to \infty }(1-\eta_i)^{n-1}=0 & \text{since no transition to~$\bot$ is taken so far}
\end{align*}
where 
$\eta_i \eqdef \min_{m \in M} \sum_{a \in A(i)}\sum_{\eta \in (0,1) \cap \Q} \, \eta\stratAct(i,m)(a_\eta)$
is a lower bound for the probability of going to $\bot$ in that step.
By the construction of $\game_\bot$, at every state $i$ and memory mode $m$,
Max chooses among actions $a_\eta$ with $a \in A(i)$ and
$\eta \in (0,1) \cap \Q$, and thus $\eta_i >0$.
\hfill\Halmos
\end{proof}

\subsection{Key Technical Lemmas}
\label{subsec-tec-lemma-trans}

The following three lemmas, \Cref{lem:key1Tr}--\cref{lem:key4Tr}, are the main technical tools, that in~\Cref{subsec:trfinal} combined with an argument  on a martingale defined on the transience values in~$\game_{\bot}$, prove \Cref{thm:Tr}. 

By \Cref{col:carrybacktoG}, all states have the same value in transience games 
$\game$ and~$\game_\bot$, and we can carry back  $\eps$-optimal memoryless  strategies  in
$\game_\bot$ to $\game$. 
Subsequently, it suffices to prove that Max has memoryless $\eps$-optimal strategies in the game $\game_\bot$.
Below, since we mostly focus on the game~$\game_{\bot}$ we may drop it from the subscripts in values, probabilities and expectations, if understood from the context. 

Without restriction, we assume that $\bot$ is the only state with value $0$
in $\game_{\bot}$, since we can redirect all transitions that
lead to a value-$0$ state to~$\bot$, without changing the value of any state.
We fix some start state~$s_0 \neq \bot$ in~$\game_{\bot}$ which
thus satisfies $\valueof{\Tr}{s_0} >0$.

Fix $u$  such that
 \begin{equation}
 \label{eq:fix-u}
 	0 \le u < \valueof{\Tr}{s_0}.
 \end{equation}

For all $n \ge 1$, let $\mathcal{F}_n$ be the sigma-algebra generated by the cylinder sets corresponding to the histories $hs \in Z^* \, S$ where $h$ starts in~$s_0$ and  visits~$s_0$ at most $n-1$ times; note that $hs$  visits $s_0$   an $n$th time if $s=s_0$ and $h$ visits~$s_0$  exactly $n-1$ times.
Let
\begin{equation}
\label{eq:tau}
	\tau:Z^{\omega}\to \{1,2,\ldots\} \cup \{\infty\}
\end{equation}
denote the random variable that counts the number of visits to $s_0$.  
Define a sequence $(X_{n})_{n=1}^{\infty}$ of random variables by
\begin{equation}
	\label{eq:defXn}
	X_{n} \ \eqdef \ \begin{cases} u & \text{if } \tau \ge n \\
                             1 & \text{if } \tau < n \text{ and } \Tr \\
                             0 & \text{if } \tau < n \text{ and not } \Tr\,.
               \end{cases}
\end{equation}
Since the event $\tau\geq n$ is the union of all cylinders generated by histories with exactly $n$ visits to $s_0$, it is 
$\mathcal{F}_n$-measurable. Moreover, restricted to the event $\tau<n$, all  
 prefixes are among the histories used to generate~$\mathcal{F}_n$; therefore any event restricted to $\tau<n$, 
  in particular $(\tau< n)\cap \Tr$, is $\mathcal{F}_n$-measurable. Hence, the $X_{n}$ are $\mathcal{F}_n$-measurable.

 
\medskip 

Denote by~$\Ret$ the event to return to the state~$s_0$, that is, the set of all plays that start in~$s_0$ and revisit it at some later step.
Intuitively, the following lemma shows that Max has a strategy that ensures
that, even if the play re-visits $s_0$, there is a positive chance of
not visiting $s_0$ again.

\begin{lemma}
\label{lem:key1Tr}
	Max has a strategy $\sigma$ such that, for all $\pi\in\ostratset$, 
	\[\probm_{s_0,\sigma,\pi}(\Tr \setminus \Ret) + \probm_{s_0,\sigma,\pi}(\Ret) u > u \,.\]
Moreover, 
if $\game_{\bot}$ is turn-based then there exists such a strategy that is deterministic.
\end{lemma}
\begin{proof}{Proof.}
Towards a contradiction suppose that 
for all Max strategies $\tilde\sigma$ there is a Min strategy $\pi$ with
\begin{equation} \label{eq:key-1}
   \probm_{s_0,\tilde\sigma,\pi}(\Tr \setminus \Ret) + \probm_{s_0,\tilde\sigma,\pi}(\Ret) u \le u.
\end{equation}
Let $\sigma$ be an arbitrary Max strategy.
For any history~${\bf h}\in H$ that ends with a visit to~$s_0$, there is a unique strategy, denoted by~$\sigma_{\bf h}$,   whose behavior is exactly like the behavior of~$\sigma$ after~${\bf h}$;
formally,  $\sigma_{\bf h}(h)=\sigma({\bf h}\,h)$ for all histories~$h\in H$. 

By~\eqref{eq:key-1}, for any history~${\bf h}$ with $s_{\bf h}=s_0$ and the unique $\sigma_{\bf h}$ just defined, there is $\pi_{\bf h}$ with 
\begin{equation}
\label{eq-pih}
	\probm_{s_0,\sigma_{\bf h},\pi_{\bf h}}(\Tr \setminus \Ret) + \probm_{s_0,\sigma_{\bf h},\pi_{\bf h}}(\Ret)\, u \, \le\,  u\,.
\end{equation}

\medskip 

Let $\pi$ be the Min strategy that upon every visit to state~$s_0$ via a history~${\bf h}$ plays $\pi_{\bf h}$ (until, if it exists, the next visit of~$s_0$).
Formally,  we define~$\pi({\bf h} \, h)=\pi_{\bf h}(h)$
if the history~${\bf h}$ ends with a visit to~$s_0$ and $h$ contains~$s_0$ only at the start. 

Consider the probability measure induced by $\game_{\bot}$ under~$\sigma$ and~$\pi$; and
recall that $X_{n}$ is $\mathcal{F}_n$-measurable. We show the following:

\begin{restatable}{claim}{claimXnsuperFn}
\label{claim:XnsuperFn}
The sequence $(X_{n})_{n=1}^{\infty}$
is a supermartingale with respect to the filtration $(\mathcal{F}_n)_{n=1}^{\infty}$:
\[\expectation[s_0,\sigma,\pi](X_{n+1} \mid \mathcal{F}_n)\leq X_{n} \,.\]
\end{restatable}
\begin{proof}
For all $n \ge 1$, write $C_n \subseteq Z^* \times \{s_0\}$ for the set of histories that end in~$s_0$ and visit~$s_0$ exactly $n$ times. For histories~$h \in C_n$, write $\Cyl{h} \subseteq Z^\omega$ for the cylinder set corresponding to~$h$.

Let $n \ge 1$ and $h \in C_n$. We first show that 
\begin{align}
\label{eq:key-2}
	\probm_{s_0,\sigma,\pi}(\Tr\text{ and } \tau =n \mid \Cyl{h})
	 + \, \probm_{s_0,\sigma,\pi}(\tau > n \mid \Cyl{h}) \cdot u \leq u
\end{align}
holds. Indeed, by the definition of~$\tau$, the plays in~$(\Tr\text{ and } \tau = n)$    
are transient plays that visit~$s_0$ exactly~$n$ times, so when starting
with~$h$, they have already visited~$s_0$ for $n$ times and will never return
to~$s_0$. Recall the relation between strategies~$\sigma$ and $\sigma_{{\bf
    h}}$ and $\pi_{{\bf h}}$ and $\pi$:

\smallskip
\begin{center}
\begin{tabular}{ c c c c}
 $\sigma$ &  $\xrightarrow{\qquad \text{\normalsize conditioned to $h$} \quad}$ & $\sigma_{{\bf h}}$ & \\ 
  &  & $\Big \downarrow$& {chosen to satisfy \eqref{eq-pih}} \\  
 $\pi$ & $\xleftarrow{\text{\normalsize reset after each visit to $s_0$} }$  & $\pi_{{\bf h}}$ &\\    
\end{tabular}
\end{center}

Similarly, the plays in~$\tau>n$ visit~$s_0$ at least~$n+1$ times, so when starting with~$h$, they must return to~$s_0$. Combining these two observations together with~\eqref{eq-pih},  we have 
\begin{equation*} 
\begin{aligned}
\probm_{s_0,\sigma,\pi}&(\Tr\text{ and } \tau = n \mid \Cyl{h}) + \probm_{s_0,\sigma,\pi}(\tau > n \mid \Cyl{h}) \cdot u \\
                       &=\ \probm_{s_0,\sigma_{{\bf h}},\pi_{{\bf h}}}(\Tr \setminus \Ret)
                         + \probm_{s_0,\sigma_{{\bf h}},\pi_{{\bf h}}}(\Ret) \cdot u && \text{by def.\ of~$\pi$}\\
&\le\ u && \text{by~\eqref{eq-pih},}
\end{aligned}
\end{equation*}
which completes the proof of~\eqref{eq:key-2}. 

\smallskip
Given an event~$E$ denote by $\mathds{1}_E$ the indicator function of~$E$.
The next step in the proof  is to show that, for all $n\geq 1$, 
\begin{align} 
\label{eq:key-2f}
\expectval_{\state_0,\zstrat,\ostrat}(\mathds{1}_{\tau\geq  n}\,  X_{n+1} \mid \mathcal{F}_n)=& \sum_{h\in C_n} \mathds{1}_{\Cyl{h}} \, \expectval_{\state_0,\zstrat,\ostrat}(X_{n+1} \mid \Cyl{h}) \,.
\end{align}
For all $n \ge 1$,	the event $\{\tau \ge n\}$ is naturally partitioned as $\{\tau \ge n\}=\biguplus_{h \in C_{n}} \Cyl{h}$. 
Hence, since $\mathds{1}_{\tau\geq  n}$ is $\mathcal{F}_n$-measurable, we have 
\begin{align*} 
\expectval_{\state_0,\zstrat,\ostrat}(\mathds{1}_{\tau\geq  n}\,  X_{n+1} \mid \mathcal{F}_n) \ 
&= \ \mathds{1}_{\tau\geq  n}\, \expectval_{\state_0,\zstrat,\ostrat}( X_{n+1} \mid \mathcal{F}_n)\\
&= \ \sum_{h\in C_n} \mathds{1}_{\Cyl{h}} \, \expectval_{\state_0,\zstrat,\ostrat}( X_{n+1} \mid \mathcal{F}_n)\,.
\end{align*}
Recall that~$\mathcal{F}_n$ is the sigma-algebra generated by the cylinder sets corresponding to the histories with at most $n$ visits to $s_0$.
Let $h\in C_n$, meaning that $h$ visits~$s_0$ for $n$ times. Since each generating cylinder set of~$\mathcal{F}_n$ either includes~$\Cyl{h}$ or has empty intersection with it, $\Cyl{h}$ is an atom in $\mathcal{F}_n$. 
The  following holds, see for example~\cite[Exercise 10, Chapter 10]{resnick1999probability}: 
\[\mathds{1}_{\Cyl{h}} \, \expectval_{\state_0,\zstrat,\ostrat}( X_{n} \mid \mathcal{F}_n)=\mathds{1}_{\Cyl{h}} \, \expectval_{\state_0,\zstrat,\ostrat}( X_{n} \mid \Cyl{h})\,,\]  
concluding the proof of~\eqref{eq:key-2f}.

Finally to prove the statement of the claim, we will show that  $\expectation[s_0,\sigma,\pi](X_{n+1} \mid \mathcal{F}_n)\leq X_{n}$ as follows. We can rewrite  $\expectation[s_0,\sigma,\pi](X_{n+1} \mid \mathcal{F}_n)$ as  $\expectation[s_0,\sigma,\pi](\mathds{1}_{\tau < n} \,   X_{n+1} + \mathds{1}_{\tau \geq  n}\,  X_{n+1} \mid \mathcal{F}_n)\,.$ But since $X_{n+1}=X_{n}$ when $\tau<n$ the above sum in turn can be written as
    \[ \expectation[s_0,\sigma,\pi]( \mathds{1}_{\tau < n} \, X_{n} \mid \mathcal{F}_n) + \, \expectation[s_0,\sigma,\pi](\mathds{1}_{\tau \geq  n}\,  X_{n+1} \mid \mathcal{F}_n) \,. \]
As $\mathds{1}_{\tau <  n}$ and $X_n$ are
  $\mathcal{F}_n$-measurable, $\expectation[s_0,\sigma,\pi]( \mathds{1}_{\tau < n} \, X_{n} \mid \mathcal{F}_n) =\mathds{1}_{\tau < n} \, X_{n}
$. 
Thus, 
  \begin{align*}
    \expectation[s_0,\sigma,\pi](X_{n+1} \mid \mathcal{F}_n)  
    &= \mathds{1}_{\tau < n} \, X_{n} + \, \expectation[s_0,\sigma,\pi](\mathds{1}_{\tau \geq  n} \, X_{n+1} \mid \mathcal{F}_n) \\ 
  &\mathop{=}^{\text{by \eqref{eq:key-2f}}} \ \mathds{1}_{\tau < n} X_{n} + \sum_{h \in C_{n}} \mathds{1}_{\Cyl{h}} \expectation[s_0,\sigma,\pi](X_{n+1} \mid \Cyl{h}) \\
  &= \ \mathds{1}_{\tau < n} \, X_{n} + \sum_{h \in C_{n}} \mathds{1}_{\Cyl{h}} \big(\probm_{s_0,\sigma,\pi}(\Tr \text{ and } \tau = n \mid \Cyl{h}) \\& \quad + \probm_{s_0,\sigma,\pi}(\tau > n \mid \Cyl{h}) \cdot u\big) \\
  &\mathop{\le}^{\text{by \eqref{eq:key-2}}} \ \mathds{1}_{\tau < n} X_{n} + \sum_{h \in C_{n}} \mathds{1}_{\Cyl{h}} u \\
  &=\  \mathds{1}_{\tau < n} \, X_{n} + \mathds{1}_{\tau \ge n} \, u \\
  &=\ \mathds{1}_{\tau < n} \, X_{n} + \mathds{1}_{\tau \ge n} \, X_{n} \\
  &= \ X_{n}\,.
\end{align*}
That is, the sequence $(X_{n})_{n=1}^{\infty}$ is a supermartingale with respect to the filtration $(\mathcal{F}_n)_{n=1}^{\infty}$.
\hfill\Halmos
\end{proof}

For all $n\geq 1$, the events $\tau+1=n$ and  $\tau+1 \leq n$ are $\mathcal{F}_n$-measurable. 
This implies that~$\tau+1$ is a stopping time with respect to the filtration~$(\mathcal{F}_n)_{n=1}^{\infty}$, given the 
 convention that $\infty + 1$ is equal~$\infty$.
Since  $\abs{X_{n}} \le 1$ for all~$n$, 
using the optional stopping theorem (OST)
\footnote{Throughout this paper we use a general version of the optional stopping theorem (OST)
from \cite[Proposition~IV-5-24, Corollary~IV-2-25]{Neveu:1975}.},
we obtain 
\begin{align*}
	u  \ 
&=   \ \expectation[s_0,\sigma,\pi] (X_{1}) \\ 
&\ge \ \expectation[s_0,\sigma,\pi] (X_{\tau+1}) && \text{by OST and \Cref{claim:XnsuperFn}}\\
&\ge \ \expectation[s_0,\sigma,\pi] (X_{\tau+1} \mid \Tr) \; \probm_{s_0,\sigma,\pi}(\Tr)\,.
\end{align*}
Since on~$\Tr$ the random variable $\tau+1 < \infty$  and $X_{\tau+1} = 1$, we  obtain  $\expectation[s_0,\sigma,\pi] (X_{\tau+1} \mid \Tr)=1$.
By the latter and the argument above, we get  
\[u \, \geq \, \probm_{s_0,\sigma,\pi}(\Tr).\]
Since the strategy~$\sigma$ was chosen arbitrarily, it follows that 
\[u \, \ge \, \valueof{\Tr}{s_0}\,,\]
a contradiction with~\eqref{eq:fix-u}.

The following claim concludes the proof of Lemma~\ref{lem:key1Tr}. 

\begin{restatable}{claim}{claimturnlemkeyoneTr}
\label{claim:turnlemkey1Tr}
If $\game_{\bot}$~is turn-based, there is a strategy~$\sigma$, as described in~\Cref{lem:key1Tr}, that is deterministic in the first step.
\end{restatable}
\begin{proof}{Proof.}
Let $|A(s_0)|>1$. Towards a contradiction, suppose that for all~$\tilde\sigma$ that are deterministic in the first step there is~$\pi$ such that $\probm_{s_0,\tilde\sigma,\pi}(\Tr \setminus \Ret) + \probm_{s_0,\tilde\sigma,\pi}(\Ret) u \le u$.
Let $\sigma$ be an arbitrary Max strategy.
It can be decomposed, in a natural way, into a distribution $\alpha \in \dist(A(s_0))$ and strategies $(\sigma_a)_{a \in A(s_0)}$, where $\sigma_a$~deterministically plays~$a$ in the first step.
By the assumption above, for each $a \in A(s_0)$ there is $\pi_a$ with $\probm_{s_0,\sigma_a,\pi_a}(\Tr \setminus \Ret) + \probm_{s_0,\sigma_a,\pi_a}(\Ret) u \le u$.
Since $\game_{\bot}$~is turn-based, we have $|B(s_0)| = 1$, and so there is a Min strategy~$\pi$ that plays $\pi_a$ for the $a \in A(s_0)$ that Max is observed to choose in the first step.
Then we have
\begin{align*}
 \probm_{s_0,\sigma,\pi}(\Tr \setminus \Ret) + \probm_{s_0,\sigma,\pi}(\Ret) u
 & \ = \ \sum_{a \in A(s_0)} \alpha(a) \left( \probm_{s_0,\sigma_a,\pi_a}(\Tr \setminus \Ret) + \probm_{s_0,\sigma_a,\pi_a}(\Ret) u \right) \\
 & \ \le \ \sum_{a \in A(s_0)} \alpha(a) u \ = \ u\,.
\end{align*}
Since $\sigma$ was arbitrary, we have derived~\eqref{eq:key-1}.
From there the proof by contradiction proceeds identically to the concurrent case.
\hfill\Halmos
\end{proof}
\end{proof}

Let $\sigma$ be as described in~\Cref{lem:key1Tr}.
Denote by~$\sigma^*$ the strategy obtained from~$\sigma$ by restarting every time $s_0$ is revisited. Formally,
 we define~
 \begin{equation}
 \label{eq-sigmastar}
 \sigma^*(h_1h_2)=\sigma(h_2)	
 \end{equation}
if the history~$h_1$ ends with a visit to~$s_0$ and $h_2$ has no visits to~$s_0$ (except at the start).
This strategy~$\sigma^*$ is not generally memoryless.
However, it takes the same mixed action whenever the play returns to~$s_0$. 
Later, when we define a memoryless strategy, we will use strategies such as~$\sigma^*$ as building blocks.


\begin{lemma}
\label{lem:key3Tr}
$\probm_{s_0,\sigma^*,\pi}(\Tr) \ge u$ for all $\pi\in \Pi$.
\end{lemma}
\begin{proof}{Proof.}
Let $\pi$ be an arbitrary Min strategy.
Consider the probability measure induced by $\game_{\bot}$ under~$\sigma^{*}$ and~$\pi$. 
 
\begin{restatable}{claim}{claimXnsubFn}
\label{claim:XnsubFn}
The sequence $(X_{n})_{n=1}^{\infty}$
is a submartingale with respect to the filtration $(\mathcal{F}_n)_{n=1}^{\infty}$:
\[\expectation[s_0,\sigma^*,\pi](X_{n+1} \mid \mathcal{F}_n)\geq X_{n} \,.\]
\end{restatable} 
\begin{proof}{Proof of the claim.}
This proof  is analogous to the proof of Claim~\ref{claim:XnsuperFn}.

As in Claim~\ref{claim:XnsuperFn}, we write $C_n \subseteq Z^* \times \{s_0\}$, with $n \ge 1$, for the set of histories that end in~$s_0$ and visit~$s_0$ exactly $n$ times. For histories~$h \in C_n$, write $\Cyl{h} \subseteq Z^\omega$ for the cylinder set corresponding to~$h$.
	
Given a history~${\bf h} \in H$ we define $\pi_{\bf h}$ for the strategy whose behavior is exactly like the behavior of~$\pi$ after~${\bf h}$.
Formally, for all histories~$h\in H$  define~$\pi_{\bf h}(h)=\pi({\bf h}\,h)$.

Let $n \ge 1$ and $h \in C_n$. Similar to~\eqref{eq:key-2} in Claim~\ref{claim:XnsuperFn}, the following can be shown:
\begin{equation} \label{eq:key-2-inv}
\begin{aligned}
&\probm_{s_0,\sigma^*,\pi}(\Tr \text{ and } \tau = n \mid \Cyl{h}) + \probm_{s_0,\sigma^*,\pi}(\tau > n \mid \Cyl{h}) \cdot u \\
&=\ \probm_{s_0,\sigma,\pi_{\bf h}}(\Tr \setminus \Ret) + \probm_{s_0,\sigma,\pi_h}(\Ret) \cdot u && \text{by definition of~$\sigma^*$ and~$\pi_{\bf h}$ }\\
&>\ u && \text{by Lemma~\ref{lem:key1Tr}.}
\end{aligned}
\end{equation}

Next, identical to~\eqref{eq:key-2f} in the proof of Claim~\ref{claim:XnsuperFn} we show that, for all $n\geq 1$, 
\begin{align} 
\label{eq:key-2finv}
\expectval_{\state_0,\zstrat^*,\ostrat}(\mathds{1}_{\tau\geq  n}\,  X_{n} \mid \mathcal{F}_n)=& \sum_{h\in C_n} \mathds{1}_{\Cyl{h}} \, \expectval_{\state_0,\zstrat^*,\ostrat}(X_{n} \mid \Cyl{h}) \,,
\end{align}
where $\mathds{1}_E$ is the indicator function of~$E$.
Finally, analogously, we  conclude that 
\begin{align*}
   \expectation[s_0,\sigma^*,\pi](X_{n+1} \mid \mathcal{F}_n)& = \mathds{1}_{\tau < n} \, X_{n} + \, \expectation[s_0,\sigma^*,\pi](\mathds{1}_{\tau \geq  n} \, X_{n+1} \mid \mathcal{F}_n) &\\ 
  &\mathop{=}^{\text{by \eqref{eq:key-2finv}}} \ \mathds{1}_{\tau < n} X_{n} + \sum_{h \in C_{n}} \mathds{1}_{\Cyl{h}} \expectation[s_0,\sigma^*,\pi](X_{n+1} \mid \Cyl{h}) \\
  &= \ \mathds{1}_{\tau < n} \, X_{n} + \sum_{h \in C_{n}} \mathds{1}_{\Cyl{h}} \big(\probm_{s_0,\sigma^*,\pi}(\Tr \text{ and } \tau = n \mid \Cyl{h}) \\
  & \quad + \probm_{s_0,\sigma^*,\pi}(\tau > n \mid \Cyl{h}) \cdot u\big) \\
  &\mathop{\ge}^{\text{by \eqref{eq:key-2-inv}}} \ \mathds{1}_{\tau < n} X_{n} + \sum_{h \in C_{n}} \mathds{1}_{\Cyl{h}} u \\
  &=\ \mathds{1}_{\tau < n} \, X_{n} + \mathds{1}_{\tau \ge n} \, u \\
  &=\ \mathds{1}_{\tau < n} \, X_{n} + \mathds{1}_{\tau \ge n} \, X_{n} \\
  &= \ X_{n}\,.
\end{align*}
That is, the sequence $(X_{n})_{n=1}^{\infty}$ is a submartingale with respect
to the filtration $(\mathcal{F}_n)_{n=1}^{\infty}$.
\hfill\Halmos
\end{proof}
 
Similarly to the proof of Lemma~\ref{lem:key1Tr}, in order to use the optional stopping theorem,  observe that~$\tau+1$ is a stopping time with respect to the filtration $(\mathcal{F}_n)_{n=1}^{\infty}$. 
Hence   
the optional stopping theorem implies that 
\begin{align*}
u \ 
&=   \ \expectation[s_0,\sigma^*,\pi](X_{1}) \\
&\le \ \expectation[s_0,\sigma^*,\pi](X_{\tau+1}) && \text{by OST and \Cref{claim:XnsubFn}}\\
&\le \ \probm_{s_0,\sigma^*,\pi}(X_{\tau+1} \ne 0)  && \text{since } 0 \le X_{\tau+1} \le 1\,.
\end{align*}
By definition of the~$X_{n}$, see~\eqref{eq:defXn},
 if $\tau < \infty$ and not~$\Tr$ then $X_{\tau+1}=0$  holds. By the above, we note that 
\begin{align*}
u \
&\le \ \probm_{s_0,\sigma^*,\pi}(X_{\tau+1} \ne 0) \\
&\le \ \probm_{s_0,\sigma^*,\pi}(\tau=\infty \text{ or } \Tr) \\
&\le \ \probm_{s_0,\sigma^*,\pi}(\tau=\infty) \, +\, \probm_{s_0,\sigma^*,\pi}(\Tr)\,.
\end{align*}

By definition of the leaky game $\game_{\bot}$ and the strategy~$\sigma^*$, there is $\delta > 0$ such that upon every visit of~$s_0$ the probability of falling into~$\bot$ in the very next step is at least~$\delta$. As a result, 
\[\probm_{s_0,\sigma^*,\pi}(\tau=\infty) = 0\]
holds, implying that $u \le \probm_{s_0,\sigma^*,\pi} (\Tr)$, as required.
\hfill\Halmos
\end{proof}

Let~$\alpha$ be a  Max mixed action at~$s_0$.
Define $\game_\alpha$ to be a game similar to~$\game_{\bot}$ where Max is only able to play $\alpha$ at~$s_0$. 
We formalize the definition of $\game_\alpha$ by introducing the notion of \emph{fixing a mixed action}~$\alpha$ in~$\game$, that is, to obtain a new game 
 identical to $\game_{\bot}$ except that 
 \begin{enumerate}
 	\item the set of   Max actions at~$s_0$ is a singleton set~$\{a_\alpha\}$; and 
 	\item for all Min actions~$b\in B(s_0)$, the transition function of $\game_\alpha$ for $(s_0,a_{\alpha},b)$ is exactly the probability distribution defined by~$\alpha$ and $b$ in~$\game_{\bot}$.
 \end{enumerate}

The following lemma shows that one can fix a mixed action at $s_0$ 
so that the value of no state drops significantly.

\begin{lemma}
\label{lem:key4Tr}
Let~$s_0 \in S$ be a state in $\game_{\bot}$ (hence $s_0 \neq \bot$) and
let $r \in (0,1)$.
There is a Max mixed action~$\alpha$
at~$s_0$ such that
\[\valueof{\game_{\alpha},\Tr}{s} \, \ge \, r \, \valueof{\game_{\bot},\Tr}{s}\]
holds for all states $s \in S$. 
If $\game_{\bot}$~is turn-based, there is such an $\alpha$ that is Dirac.
\end{lemma}
\begin{proof}{Proof.}
Let $s \in S$ be chosen arbitrarily.
Let 
\begin{align}
\label{def-new-u}
	u \eqdef r \, \valueof{\game_{\bot},\Tr}{s_0}
\end{align}
ensuring that $u < \valueof{\game_{\bot},\Tr}{s_0}$ as required in~\eqref{eq:fix-u}.
Let $\sigma$ be a strategy satisfying the condition in~\Cref{lem:key1Tr} for this~$u$. From~$\sigma$, we obtain~$\sigma^*$ as described in~\eqref{eq-sigmastar}.
Let the mixed action~$\alpha$, a probability distribution  over~$A(s_0)$, be
the action of~$\sigma^*$ at history~$s_0$, i.e.,
\begin{align}
\label{def-alpha-sigmastar}
\alpha \eqdef \sigma^*(s_0)
\end{align}
If $\game_{\bot}$~is turn-based, $\alpha$~is Dirac.

Let  $x \in (0,1)$ be chosen arbitrarily. 
We finally define~$\tilde{\sigma}$ from~$\sigma^*$:
Starting in~$s$,  this strategy~$\tilde{\sigma}$ first plays a strategy~$\sigma_s$ with 
\begin{equation}
\label{eq:defsigmas}
	\inf_{\pi} \, \probm_{\game_{\bot},s,\sigma_s,\pi}(\Tr) \ge x \, \valueof{\game_{\bot},\Tr}{s}\,,
\end{equation} and 
whenever (if at all) the play visits~$s_0$, it switches to~$\sigma^*$. 

Let $\tilde{\pi}$~be an arbitrary Min strategy.
We redefine the random variable~$\tau$ as given in~\eqref{eq:tau},
allowing for the possibility that $\tau = 0$ to accommodate plays that never visit~$s_0$. 
For the~$u$ in~\eqref{def-new-u}, define
 the  sequence $(X_{n})_{n=1}^{\infty}$ of random variables as in~\eqref{eq:defXn}.
Define 
\begin{equation}
\label{eq:defX0}
	X_{0} \eqdef r \, x\, \valueof{\game_{\bot},\Tr}{s}\,,
\end{equation}
and let $\mathcal{F}_0$ be the sigma-algebra generated by the cylinder set
corresponding to the history that consists only of the initial state~$s$. 
Similar to the $X_{n}$ being $\mathcal{F}_n$-measurable,
the random variable~$X_{0}$ is also $\mathcal{F}_0$-measurable. 

\begin{restatable}{claim}{claimXnzerosubFn}
\label{claim:Xn0subFn}
The sequence $(X_{n})_{n=0}^{\infty}$
is a submartingale with respect to the filtration $(\mathcal{F}_n)_{n=0}^{\infty}$:
\[\expectation[\game_{\bot},s_0,\tilde{\sigma},\tilde{\pi}](X_{n+1} \mid \mathcal{F}_n)\geq X_{n} \,.\]
\end{restatable}
\begin{proof}{Proof of the claim.}
The proof of the this Claim is a combination of
reasoning for $\expectation[\game_{\bot},s,\tilde{\sigma},\tilde{\pi}](X_{1}
\mid \mathcal{F}_0) \geq X_{0}$ and Claim~\ref{claim:XnsubFn}.
In this proof all probabilities and expectations refer to the game $\game_\bot$.  
Denote by~$\Vis$ the event that $s_0$~is visited. Following the definition of~$X_{1}$ in~\eqref{eq:defXn}:  
\begin{align*}
\expectation[s,\tilde{\sigma},\tilde{\pi}](X_{1} \mid \mathcal{F}_0) \
&= \ \probm_{s,\sigma_s,\tilde{\pi}}(\Vis) \cdot u + \probm_{s,\sigma_s,\tilde{\pi}}(\Tr\text{ and not } \Vis) \cdot 1 \\& +\, \probm_{s,\sigma_s,\tilde{\pi}}(\text{not } \Tr \text{ and not }\Vis) \cdot 0 \\
\intertext{Substituting~$u$ from~\eqref{def-new-u}  we obtain } 
&=\ r \, \probm_{s,\sigma_s,\tilde{\pi}}(\Vis) \valueof{\Tr}{s_0} + \probm_{s,\sigma_s,\tilde{\pi}}(\Tr \text{ and not } \Vis) \\
&\ge\ r \, \big( \probm_{s,\sigma_s,\tilde{\pi}}(\Vis) \valueof{\Tr}{s_0} + \probm_{s,\sigma_s,\tilde{\pi}}(\Tr \text{ and not } \Vis) \big) \\
&\ge\ r \inf_\pi \big( \probm_{s,\sigma_s,\pi}(\Vis) \valueof{\Tr}{s_0} + \probm_{s,\sigma_s,\pi}(\Tr \text{ and not } \Vis) \big) \\
&\ge\ r \inf_\pi \probm_{s,\sigma_s,\pi}(\Tr) \\
\intertext{By \eqref{eq:defsigmas}, we further have that}
&\ge\ r \, x \, \valueof{\Tr}{s} \\
&=\ X_{0}\,.
\end{align*}

Now  the proof given for the Claim~\ref{claim:XnsubFn} showing that 
$(X_{n})_{n=1}^{\infty}$ is a submartingale with respect to the filtration $(\mathcal{F}_n)_{n=1}^{\infty}$, combined with the calculation above concludes the statement of the Claim.
\hfill\Halmos
\end{proof}

Proceeding similarly as in the proof of \cref{lem:key3Tr}, we can get that
\begin{align*}
r \, x \, \valueof{\game_{\bot},\Tr}{s} \ 
  &=   \ \expectation[\game_{\bot},s,\tilde{\sigma},\tilde{\pi}](X_{0}) &&
\text{by definition in \eqref{eq:defX0} and $X_0$ constant}\\
&\le \ \expectation[\game_{\bot},s,\tilde{\sigma},\tilde{\pi}](X_{\tau+1}) && \text{by OST and \Cref{claim:Xn0subFn}} \\
&\le \ \probm_{\game_{\bot},s,\tilde{\sigma},\tilde{\pi}}(\Tr) &&\text{similarly to Lemma~\ref{lem:key3Tr}.}
\end{align*}

By its construction, $\tilde{\sigma}$ chooses the mixed action
$\alpha$ from \eqref{def-alpha-sigmastar}
each time the play visits~$s_0$.
Recall that $\game_{\alpha}$ is the game
obtained from~$\game_{\bot}$ by fixing~$\alpha$ at~$s_0$, ensuring that Max is only able to play $\alpha$ at~$s_0$.
Hence, the  strategy $\tilde{\sigma}$  is uniquely translated to a corresponding strategy~$\tilde{\sigma}_{\alpha}$ in $\game_{\alpha}$ in a natural way.   
Following 
$\tilde{\sigma}_{\alpha}$  in~$\game_{\alpha}$,  we obtain 
\[r \, x \, \valueof{\game_{\bot},\Tr}{s} \le \probm_{\game_{\alpha},s,\tilde{\sigma}_{\alpha},\tilde{\pi}}(\Tr).\]
Since $x$ and~$\tilde{\pi}$ were arbitrary, it follows that 
\[r \, \valueof{\game_{\bot},\Tr}{s} \le \valueof{\game_{\alpha},\Tr}{s}\,,\] as required.
\hfill\Halmos
\end{proof}

\subsection{Proof of \Cref{thm:Tr}}
\label{subsec:trfinal}
We are now in a position to 
prove \Cref{thm:Tr} which states that Max has memoryless strategies that are $\eps$-optimal from every state in a transience game. 

In the sequel, for clarity, we assume that the set of states in~$\game_{\bot}$, excluding~$\bot$,  is $\{1,2,\ldots\}$.
Let~$r \in (0,1)$ be arbitrary. 
Define
\(\game_0 \eqdef \game_{\bot}\). We inductively define a sequence  $(\game_i)_{i\geq 1}$ of games and a sequence $(\alpha_i)_{i\geq 1}$ of Max mixed actions. 
For all~$i\geq 1$, invoking
 Lemma~\ref{lem:key4Tr}, there exists a probability distribution~$\alpha_i$ on~$A(i)$  such that   
for all $s \in S$
 \begin{equation}
\label{eq:defGi}
	\valueof{\game_i,\Tr}{s} \ge r^{2^{-i}} \, \valueof{\game_{i-1},\Tr}{s}\,
\end{equation}
where the game $\game_i$ is obtained from $\game_{i-1}$ by fixing the mixed actions~$\alpha_i$ at~$i$.

The collection of the  mixed actions~$\alpha_i$ at states~$i$ defines a memoryless strategy, denoted by~$\sigma$, for Max in $\game_{\bot}$. 
By \Cref{lem:key4Tr}, if $\game_{\bot}$~is turn-based, $\sigma$~can be made deterministic. Below, we will prove that, for all states~$s \in S$ and for all Min strategies~$\pi$, 
\begin{equation}
\label{eq:probGTr}
	\probm_{\game_{\bot},s,\sigma,\pi}(\Tr) \ge r \, \valueof{\game_{\bot},\Tr}{s}\,,
\end{equation}
proving \Cref{thm:Tr}.

For all $i \ge 0$,  define $r_i \eqdef \prod_{j>i} r^{2^{-j}}$, that is intuitively the  proportion of the value in the game~$\game_i$ that will be maintained by all  \emph{future fixings}~$j>i$ .
For brevity,  write $v_i(s)$ for $\valueof{\game_i,\Tr}{s}$.
It follows from the definition of the games~$\game_i$ in \eqref{eq:defGi}
 that for all $i \ge 1$ and all states~$s$,
 \begin{equation} \label{eq:r-value-inv}
 r_{i} \, v_i(s) \ \ge \ r_{i-1} \, v_{i-1}(s) \,.
\end{equation}

Recall that $H_{n}$~is the set of histories at step $n$.
These histories visit $n+1$ states (though not necessarily $n+1$ different states).
For all $n \ge 1$, let $\mathcal{F}_n$ be the sigma-algebra generated by the cylinder sets corresponding to~$H_{n-1}$.

Let $\pi$ be an arbitrary Min strategy.
Given a play visiting the sequence $s_1, s_2, \ldots $ of states, for all $n \ge 1$,  we define the random variable
\[m(n)\eqdef \max\{s_1,s_2,\ldots,s_n\}.\]
 We further define 
the random variable~$Y_n$, taking values in~$[0,1]$, by 
\begin{equation}
\label{eq:defYn}
	Y_n \eqdef r_{m(n)} \, v_{m(n)}(s_n)  \,.
\end{equation}
Intuitively, the variable $Y_n$ is  the value  
 of~$s_n$ after having \emph{fixed} an action in the states $\{1,\cdots,m(n)\}$, scaled down by  $r_{m(n)}$, that is the proportion of the value  in the game~$\game_i$ that will be maintained by all  \emph{future fixings}~$i>m(n)$.
Both random variables~$m(n)$ and~$Y_n$ are $\mathcal{F}_n$-measurable. 

Let $s_1$ be an arbitrary state, excluding~$\bot$. We aim to prove the inequality in~\eqref{eq:probGTr}.  
Consider a random play visiting the sequence $s_1, s_2, \ldots $ of states.
For all $n \ge 1$,  we have
\begin{align*}
\expectation[s_1,\sigma,\pi](Y_{n+1} \mid \mathcal{F}_n) \
&   = \ \expectation[s_1,\sigma,\pi](r_{m(n+1)} v_{m(n+1)}(s_{n+1}) \mid \mathcal{F}_n) & \text{by } \eqref{eq:defYn} \\
& \ge \ \expectation[s_1,\sigma,\pi](r_{m(n)} v_{m(n)}(s_{n+1}) \mid \mathcal{F}_n) & \text{by~\eqref{eq:r-value-inv}} \\
\intertext{But then recall that in state~$s_n$, strategy~$\sigma$ plays the mixed action  fixed in~$\game_{m(n)}$, and, therefore, the expected value  of the values in~$\game_{m(n)}$ does not drop in one step. Hence, we get }
& \ge \ \expectation[s_1,\sigma,\pi](r_{m(n)} v_{m(n)}(s_n) \mid \mathcal{F}_n) & \\
&   = \ Y_n\;.
\end{align*}
Consequently, the sequence $(Y_n)_{n=1}^{\infty}$
is a submartingale with respect to the filtration $(\mathcal{F}_n)_{n=1}^{\infty}$. This implies that, in game~$\game_{\bot}$, for all $n \geq 1$,  
\begin{equation} \label{eq:submart}
\expectation[s_1,\sigma,\pi](Y_{n}) \ \ge \ \expectation[s_1,\sigma,\pi](Y_{1}). 
\end{equation}
We conclude the proof as follows. In game~$\game_{\bot}$, by \Cref{lem:strongerTrNotbot} and $\sigma$ being memoryless, we get that  
\begin{align*}
\probm_{s_1,\sigma,\pi}(\Tr) \
   &= \ \probm_{s_1,\sigma,\pi}(\avoid[\bot])   \\
\intertext{Since $\bot$ is a sink state and  has value~$0$ for transience,  we have}
& \ge \ \probm_{s_1,\sigma,\pi}\Big(\limsup_{n \to \infty} Y_n > 0\Big) & \\
\intertext{Furthermore, since \text{$Y_n \le 1$}, it follows that}
& \ge \ \expectation[s_1,\sigma,\pi]\Big(\limsup_{n \to \infty} Y_n\Big) &  \\
& \ge \ \limsup_{n \to \infty}~ \expectation[s_1,\sigma,\pi] (Y_n) && \text{reverse Fatou lemma} \\
& \ge \ \expectation[s_1,\sigma,\pi] (Y_{1}) && \text{by~\eqref{eq:submart}} \\
&   = \ r_1 \, v_1(s_1) && \text{by~\eqref{eq:defYn}} \\
& \ge \ r_0 \, v_0(s_1) && \text{by~\eqref{eq:r-value-inv}}
\intertext{By~\eqref{eq:defGi} and the definition of~$r_0$, we finally can conclude that}
\probm_{s_1,\sigma,\pi}(\Tr) \ 
&\geq \ r \, \valueof{\game_{\bot},\Tr}{s_1}\,,
\end{align*}
as required in~\eqref{eq:probGTr}.

\section{B\"uchi Games}
\label{sec-buchi}
In a game $\game$ with state space~$S$ and $T \subseteq S$ denote by~$\Bu$ the
B\"uchi objective,
i.e., the objective to visit the set $T$ infinitely often.
(If the set $T$ is infinite then some plays that satisfy $\Bu$
might not visit any particular state in $T$ infinitely often.)
Define also the transient B\"uchi objective, $\TB \eqdef \transient \cap \Bu$.
Clearly, $\TB$~can only be met if $T$~is infinite.
We will prove the following main result.

\thmTrBuchi*

We prove \cref{thm:TrBuchi} in \cref{sub:proof-thm-TrBuchi}.
Then, in \cref{sub:proof-thm-Buchi} we show that \Cref{thm:TrBuchi} implies \Cref{thm:Buchi}, as announced in \Cref{sec-overview}.

\subsection{Proof of \Cref{thm:TrBuchi}} \label{sub:proof-thm-TrBuchi}

Fix a game $\game$, and consider the derived leaky game $\game_\bot$ as defined in \Cref{sec-trans}.
By \cref{col:carrybacktoG}, all states have the same value in
$\game$ and $\game_\bot$ wrt.\ the $\TB$ objective, and we can carry back  $\eps$-optimal $1$-bit  strategies  in
$\game_\bot$ to $\game$. 
Thus, it suffices to prove that Max has $\eps$-optimal $1$-bit strategies in
the game $\game_\bot$.

We will assume without loss of generality that $\bot\notin S$~is the only
state with $\valueof{\TB}{\bot} = 0$, as we can redirect all transitions that
lead to a value-$0$ state to~$\bot$, without changing the value of any state.
Moreover, by \Cref{lem:strongerTrNotbot}, in~$\game_\bot$,
under every finite-memory Max strategy
and arbitrary Min strategy,
the events $\transient$ and~$\avoid[\bot]$ are equal up to measure zero.

Intuitively, the \emph{bridge} in the following definition fixes a strategy
in a subspace $I \cup L$.
One starts at a set of initial states $I$, continues throughout
the set of states $L$ and finally exits the subspace at some state in the
target set $T$.

\begin{definition}\label{def:bridge}
A \emph{bridge} in~$\game_\bot$ is a triple $\Lambda = (I,L,\sigma_0)$ where
$I, L \subseteq S$ are disjoint finite sets of states 
and $\sigma_0$ is a mapping that
maps every $s \in I \cup L$ to a mixed Max action $\sigma_0(s) \in \dist(A(s))$.
(Since we assume that $\bot \notin S$, we also have $\bot \notin I,L,T$).
We write $\bev{\Lambda}$ for the event that after each visit of a state in~$I$, eventually $I \cup L$ will be left, at which point immediately a state in~$T$ is visited.

A Max strategy~$\sigma$ is said to be \emph{consistent} with~$\Lambda$ if,
in any play against any Min strategy $\pi$,
after every visit of a state in~$I$, strategy~$\sigma$ plays as specified by~$\sigma_0$ until $I \cup L$~is left.
A bridge is called \emph{deterministic} if,
for every $s \in I \cup L$,
$\sigma_0(s)$ is Dirac,
i.e., it chooses a pure action.
\end{definition}

To prove \Cref{thm:TrBuchi}, we need the following key lemma,
whose role is similar to the role of \Cref{lem:key4Tr} in \Cref{sec-trans}.
  
\begin{restatable}{lemma}{lembukey}\label{lem:Bu-key}
  Let $I \subseteq S$ be a finite set of states in $\game_\bot$ 
  (thus $\bot\notin I$). Let $r \in (0,1)$.
There are a bridge $\Lambda = (I,L,\sigma_0)$ and a Max strategy~$\sigma$
consistent with~$\Lambda$
so that 
\[
  \inf_\pi\probm_{s,\sigma,\pi}(\bev{\Lambda} \cap \TB) \ge r \,
  \valueof{\TB}{s} 
\]
holds for all states~$s \in S$.
If $\game_\bot$ is turn-based then there is a deterministic such bridge.
\end{restatable}

For the proof of \Cref{lem:Bu-key}, we assume that 
all measures and expectations
are w.r.t.\ the leaky game $\game_\bot$.

First we will need the following general technical lemma.

\begin{lemma} \label{lem:technical}
Let $f : (0,1]^n \to (0,1]^n$ be any continuous function.
For any $x \in (0,1)^n$ there is a $y \ge x\in (0,1)^n$ such that
for all $i \in \{1, \ldots, n\}$
\[
f(y)_i > y_i \ \Leftrightarrow\ y_i > x_i\,.
\]  
\end{lemma}
\begin{proof}{Proof.}
We need to show that 
for any $x \in (0,1)^n$ there is a $y \ge x \in (0,1)^n$ such that for all $i \in \{1, \ldots, n\}$
\[
  f(y)_i > y_i \ \Longrightarrow \ y_i > x_i \quad \text{ and } \quad f(y)_i \le y_i \ \Longrightarrow \ y_i = x_i\,.
\]
Define $y^{(1)} \eqdef x$.
For any $m \ge 1$ for which $y^{(m)}$ has been defined, let
\[
E^{(m)} \ \eqdef \ \{i \in \{1, \ldots, n\} \mid f(y^{(m)})_i > y^{(m)}_i = x_i \}\,.
\]
If $E^{(m)} = \emptyset$, define $y \eqdef y^{(m)}$.
Otherwise, since $f$~is continuous, we can define a vector $y^{(m+1)}$
with $y^{(m+1)} \ge y^{(m)}$ (in all components) such that $1 \ge f(y^{(m+1)})_i > y^{(m+1)}_i > y^{(m)}_i$ holds for all $i \in \{1, \ldots, n\}$ with $f(y^{(m)})_i > y^{(m)}_i$.
Since the $E^{(m)}$ are pairwise disjoint, this defines the required vector~$y$.
\hfill\Halmos
\end{proof}

Fix a finite set of states $I \subseteq S$ with $\bot \notin I$.
For $s \in I$, define $\Ret_s$ as the following event: visit a state in
$T \setminus I$; then (necessarily at least one step later)
visit a state in~$I$ and the first such visited state
in~$I$ is~$s$.
Events $(\Ret_s)_{s \in I}$ are pairwise disjoint.
Define $\Ret \eqdef \bigcup_{s \in I} \Ret_s$.
Towards a proof of \cref{lem:Bu-key} we first prove the following lemma.
(Here and in similar cases, we often write $u_s$ for $u(s)$, where $u \in (0,1]^I$ and $s \in I$.)

\begin{lemma} \label{lem:Bu-key-aux}
Let $q \in (0,1)$.
There are $u \in (0,1)^{I}$
with $u_s \ge q \valueof{\TB}{s}$ for all $s \in I$
and for all $s_0 \in I$ there is
a Max strategy~$\sigma$ such that
\[
 \inf_\pi \, \Big( \probm_{s_0,\sigma,\pi}(\TB \setminus \Ret) + \sum_{s \in I} \probm_{s_0,\sigma,\pi}(\Ret_s) u_s \Big) \ > \ u_{s_0}
\]
\end{lemma}
\begin{proof}{Proof.}
Recall that we assume without loss of generality that $\bot \notin I$ and that $\bot$~is the only state
with $\valueof{\TB}{\bot} = 0$.
Consider the function $f : (0,1]^{I} \to \mathbb{R}^{I}$ defined by,
for all $s_0 \in I$,
\[
 f(x)_{s_0} \ \eqdef \ \sup_\sigma \, \inf_\pi \, \Big(
 \probm_{s_0,\sigma,\pi}(\TB \setminus \Ret) + \sum_{s \in I}
 \probm_{s_0,\sigma,\pi}(\Ret_s) x_s \Big)
\]
Intuitively, one may view $f(x)_{s_0}$ as the (lower) value of~$s_0$ with respect to a particular objective where the ``reward'' upon~$\Ret$ is given by~$x$.
From the assumptions on $I$ and the
disjointness of the events $(\Ret_s)_{s \in I}$, it follows
that $f$ is also $f : (0,1]^{I} \to (0,1]^{I}$.
Since $\probm_{s_0,\sigma,\pi}(\Ret_s) \in [0,1]$ is bounded,
$f$~is continuous.
Thus we can use \cref{lem:technical}, instantiated with
$n = |I|$ and
$x_{s_0} = q \valueof{\TB}{s_0} \in (0,1)$,
and obtain that
there is $u \in (0,1)^{I}$ such that for all $s_0 \in I$ we have $u_{s_0} \ge q \valueof{\TB}{s_0}$, and we have $f(u)_{s_0} > u_{s_0}$ if and only if $u_{s_0} > q \valueof{\TB}{s_0}$.
Let $J \eqdef \{s_0 \in I \mid u_{s_0} = q \valueof{\TB}{s_0}\}$.
It suffices to show that $J = \emptyset$.

Choose $q' \in (q,1)$ such that $u_{s} > q' \valueof{\TB}{s}$ for all $s \in I \setminus J$.
For $s_0 \in J$ and a strategy~$\sigma$ write
\[
 g(s_0,\sigma) \ \eqdef \ \inf_\pi \, \Big(\probm_{s_0,\sigma,\pi}(\TB \setminus \Ret) + \sum_{s \in I} \probm_{s_0,\sigma,\pi}(\Ret_s) u_s\Big) \,.
\]
Intuitively, one may view $g(s_0,\sigma)$ as the attainment of~$\sigma$ starting in~$s_0$ with respect to the particular objective mentioned above in the context of~$f$; however, $x$ is now fixed to be~$u$.
It follows from the properties of~$u$ and the definition of~$J$ that $f(u)_{s_0} \le u_{s_0}$ holds for all $s_0 \in J$.
Thus, for all $s_0 \in J$ and all~$\sigma$ we have $g(s_0,\sigma) \le u_{s_0}$.
It follows from the construction of~$\game_\bot$ that there is always
a \emph{strictly} ``better'' strategy~$\sigma'$, i.e., $g(s_0,\sigma') > g(s_0,\sigma)$.
Thus, for all $s_0 \in J$ and all~$\sigma$ we have $g(s_0,\sigma) < u_{s_0}$.
Hence,
\begin{equation} \label{eq:lem-Bu-key-aux-0}
\forall\, s_0 \in J \ \forall\, \sigma \ \exists\, \pi: \  \probm_{s_0,\sigma,\pi}(\TB \setminus \Ret) + \sum_{s \in I} \probm_{s_0,\sigma,\pi}(\Ret_s) u_s \ \le \ u_{s_0} \,.
\end{equation}
(We could even have $< u_{s_0}$ above, but that is not needed.)
For a play starting from a state in~$I$, we define in the following a sequence of states, namely $s^{(1)}, s^{(2)}, \ldots \in I$.
This sequence may be finite, in which case $s^{(n)}$~is no longer defined from some $n$ on.
Each defined~$s^{(n)}$ naturally corresponds to a visit of~$I$.
Specifically, denote by $s^{(1)} \in I$ the initial state, and, for $n \ge 1$, denote by $s^{(n+1)} \in I$ the state for which the play suffix started from~$s^{(n)}$ satisfies $\Ret_{s^{(n+1)}}$.
In other words, after the visit of~$s^{(n)}$ a state in $T \setminus I$ is visited and then a state in~$I$ is visited and the first such visited state is~$s^{(n+1)}$.
Denote by~$\tau$ the random variable, taking values in $\{1,2,\ldots\} \cup \{\infty\}$, such that for all $n \in \{1, 2, \ldots\}$ we have that $s^{(n)}$~is defined if and only if $\tau \ge n$.
Intuitively, $\tau$~is the number of visits of~$I$ with visits of $T \setminus I$ in between.
For all $n \in \{1,2,\ldots\}$, let $\mathcal{F}_n$ be the sigma-algebra generated by the cylinder sets corresponding to the histories $w s \in Z^* \times S$ such that $w$~includes at most the visits $s^{(1)}, \ldots, s^{(n-1)}$ (i.e., the state~$s$ at the end of history~$w s$ may be the visit $s^{(n)}$).
Note that the event $\tau \ge n$ and the random states $s^{(1)}, \ldots, s^{(\min\{\tau,n\})}$ are $\mathcal{F}_n$-measurable.

Let $\tau'$ be the random variable, taking values in $\{1, 2, \ldots\} \cup \{\infty\}$, such that if there is $n \in \{1, 2, \ldots\}$ with $s^{(1)}, \ldots, s^{(n-1)} \in J$ and $s^{(n)} \in I \setminus J$ then $\tau' \eqdef n$, and $\tau' \eqdef \tau+1$ if such~$n$ does not exist.
Note that \emph{either} $\infty > \tau' \le \tau$ and $s^{(\tau')} \in I \setminus J$ is defined 
\emph{or} $\tau' = \tau + 1$ and $s^{(\tau')}$~is not defined. 
Random variable~$\tau'$ is a stopping time with respect to the filtration $\mathcal{F}_1, \mathcal{F}_2, \ldots$.
Define random variables $X_1, X_2, \ldots$ by
\[
X_n \ \eqdef \ \begin{cases} u_{s^{(n)}}     & \text{if } \tau' > n \\
                             u_{s^{(\tau')}} & \text{if } \tau' \le n \text{ and } \tau' \le \tau \\
                             1 & \text{if } \tau' = \tau + 1 \le n \text{ and } \TB \\
                             0 & \text{if } \tau' = \tau + 1 \le n \text{ and not } \TB\,.
               \end{cases}
\]
Each $X_n$ is $\mathcal{F}_n$-measurable.

Let $\sigma$ be an arbitrary strategy.
The behavior of~$\sigma$ after any visit of~$I$ is determined by the history leading up to the visit.
Thus, for every history~$h$ that ends with a visit of~$I$ there is a strategy, say $\sigma_h$, whose behavior is exactly like the behavior of~$\sigma$ after~$h$.
By~\eqref{eq:lem-Bu-key-aux-0} and the choice of~$q'$, for every $s_0 \in I$
and every~$h$ that ends with a visit of~$s_0$ there exists a (sufficiently
good) Min strategy~$\pi_h$ such that\\
if $s_0 \in J$ then
\begin{equation}\label{eq:sigma-h1}
\probm_{s_0,\sigma_h,\pi_h}(\TB \setminus \Ret) + \sum_{s \in I} \probm_{s_0,\sigma_h,\pi_h}(\Ret_s) u_s \le u_{s_0}
\end{equation}
and if $s_0 \in I \setminus J$ then (since $u_{s_0} > q' \valueof{\TB}{s_0}$)
\begin{equation}\label{eq:sigma-h2}
u_{s_0} \ge q' \probm_{s_0,\sigma_h,\pi_h}(\TB).
\end{equation}
Let $\pi$ be the Min strategy that upon every visit~$s^{(n)}$ with $n \le \min\{\tau,\tau'\}$ plays~$\pi_h$ (until the visit~$s^{(n+1)}$ with $n+1 \le \min\{\tau,\tau'\}$ if this visit exists, and forever otherwise).

For all $n \ge 1$ and $s_0 \in J$, write $C_{n,s_0} \subseteq Z^* \times \{s_0\}$ for the set of histories that include the visits $s^{(1)}, \ldots, s^{(n)} \in J$ and end with $s^{(n)} = s_0$.
For any $h \in C_{n,s_0}$, write $\Cyl{h} \subseteq Z^\omega$ for the cylinder set corresponding to~$h$.
For all $n \ge 1$ we have $\{\tau' > n\} = \bigcup_{s_0 \in J} \bigcup_{h \in C_{n,s_0}} \Cyl{h}$, where the union is disjoint.
From the definitions of $\pi$ and~$\pi_h$
(\Cref{eq:sigma-h1,eq:sigma-h2})
it follows that for all $\tilde s \in J$ and $n \ge 1$ and $s_0 \in J$ and $h \in C_{n,s_0}$
\begin{equation} \label{eq:lem-Bu-key-aux-1}
\begin{aligned}
\expectation[\tilde s,\sigma,\pi](X_{n+1} \mid \Cyl{h}) \
&=\ \probm_{\tilde s,\sigma,\pi}(\tau'-1 = \tau = n \text{ and } \TB \mid \Cyl{h}) \\
& \hspace{5mm} \mbox{} + \sum_{s \in I} \probm_{\tilde s,\sigma,\pi}(\tau \ge n+1 \text{ and } s^{(n+1)}=s \mid \Cyl{h}) u_s \\
&=\ \probm_{s_0,\sigma_h,\pi_h}(\TB \setminus \Ret) + \sum_{s \in I} \probm_{s_0,\sigma_h,\pi_h}(\Ret_s) u_s \\
&\le\ u_{s_0}\,.
\end{aligned}
\end{equation}

Towards a contradiction, suppose that $J \ne \emptyset$, i.e., there is $\tilde s \in J$.
To avoid clutter, let us drop the subscripts from $\probm_{\tilde{s},\sigma,\pi}$ and $\expectation[\tilde{s},\sigma,\pi]$ in the following. For all $n \ge 1$
\begin{align*}
 \expectation(X_{n+1} \mid \mathcal{F}_n) \
  &= \ \mathds{1}_{\tau' \le n} X_n + \sum_{s_0 \in J} \sum_{h \in C_{n,s_0}} \mathds{1}_{\Cyl{h}} \expectation(X_{n+1} \mid \Cyl{h}) \\
  &\mathop{\le}^{\text{\eqref{eq:lem-Bu-key-aux-1}}} \ \mathds{1}_{\tau' \le n} X_n + \sum_{s_0 \in J} \sum_{h \in C_{n,s_0}} \mathds{1}_{\Cyl{h}} u_{s_0} \\
  &=\ \mathds{1}_{\tau' \le n} X_n + \sum_{s_0 \in J} \mathds{1}_{\tau'>n,\ s^{(n)}=s_0} u_{s_0} \\
  &=\ \mathds{1}_{\tau' \le n} X_n + \mathds{1}_{\tau'>n} u_{s^{(n)}} \\
  &=\ \mathds{1}_{\tau' \le n} X_n + \mathds{1}_{\tau'>n} X_n \\
  &= \ X_n\,,
\end{align*}
where $\mathds{1}_E$ is the indicator function of~$E$.
That is, $X_1, X_2, \ldots$ is a supermartingale with respect to the filtration $\mathcal{F}_1, \mathcal{F}_2, \ldots$.
Using the optional stopping theorem (noting that $|X_n| \le \max\{1,\max_{s
  \in I} u_s\} \le 1$ for all~$n$), we obtain
\begin{align}
q \valueof{\TB}{\tilde s} \ &= u_{\tilde s} \ = \ \expectation X_1 \ \ge \ \expectation X_{\tau'} \notag \\
                            &\ge \probm(\tau' = \tau+1 \wedge \TB) \notag\\
                                & \quad + \probm(\infty > \tau' \le \tau) \expectation (u_{s^{(\tau')}} \mid \infty > \tau' \le \tau)\,. \label{eq:lem-Bu-key-aux-2}
\end{align}
Considering the second summand of the sum above, we have
\begin{align*}
&\probm(\infty > \tau' \le \tau) \expectation (u_{s^{(\tau')}} \mid \infty > \tau' \le \tau) \\
&=\ \sum_{s \in I \setminus J} \probm\big(\infty > \tau' \le \tau \wedge s^{(\tau')}=s\big) u_s \\
&\ge\ q' \sum_{s \in I \setminus J} \probm\big(\infty > \tau' \le \tau \wedge s^{(\tau')}=s\big) \probm(\TB \mid \infty > \tau' \le \tau \wedge s^{(\tau')}=s) \\
&=\ q' \sum_{s \in I \setminus J} \probm\big(\infty > \tau' \le \tau \wedge s^{(\tau')}=s \wedge \TB\big) \\
&= \  q' \probm(\infty > \tau' \le \tau \wedge \TB)\,,
\end{align*}
where the inequality follows from the definition of~$\pi$.
By combining this with~\eqref{eq:lem-Bu-key-aux-2} we obtain
\begin{align*}
   q \valueof{\TB}{\tilde s} \
  &\ge \ q' \big(\probm(\tau' = \tau+1 \text{ and } \TB) + \probm(\infty > \tau' \le \tau \text{ and } \TB)\big) \\
  & = \ q' \probm_{\tilde{s},\sigma,\pi}(\TB)\,.
\end{align*}
Since the strategy~$\sigma$ was chosen arbitrarily, it follows that
$q \valueof{\TB}{\tilde s} \ge q' \valueof{\TB}{\tilde s}$.
Since $\valueof{\TB}{\tilde s}>0$, we obtain $q \ge q'$, a contradiction.
\hfill\Halmos
\end{proof}

For $s \in I$, define $\Vis_s$ as the event that a state in~$I$ is visited and the first such visited state is~$s$.
Events $(\Vis_s)_{s \in I}$ are pairwise disjoint.
Define $\Vis \eqdef \bigcup_{s \in I} \Vis_s$.

Let $u \in (0,1)^{I}$ be as in \cref{lem:Bu-key-aux}.
Extend $u$ so that $u \in (0,1]^{I \cup T}$ by defining,
for all $f \in T$,
\begin{equation}\label{eq:def-uf}
 u_f \ \eqdef \ \sup_{\sigma} \, \inf_\pi \, \Big( \probm_{f,\sigma,\pi}(\TB \setminus \Vis) + \sum_{s \in I} \probm_{f,\sigma,\pi}(\Vis_s) u_s \Big). 
\end{equation}
We may have $u_f=1$ for $f \in T\setminus I$.
Since $I \cap T$ may be non-empty, the definition of $u_f$ above
for $u_f \in I \cap T$ might conflict with the construction of $u$ from
\cref{lem:Bu-key-aux}.
However, for $u_f \in I \cap T$ the definition above reduces to $u_f = u_f$,
since $f$ is the first state in $I$ that is visited (immediately),
i.e., $\probm_{f,\sigma,\pi}(\Vis_f) = 1$ holds for all $f \in I \cap T$ and all~$\sigma,\pi$.
Thus the extended $u$ is well-defined. 

Let $U$~denote the random variable such that $U = u_f$ where $f$~is the first visited state in $T \setminus I$, and $U = 0$ if $T \setminus I$ is not visited.
Next we prove the following lemma.
\begin{lemma} \label{lem:Bu-key-aux-1}
Let $\sigma$ be the strategy from \cref{lem:Bu-key-aux}.
We have
$
 \inf_\pi\, \expectation[s_0,\sigma,\pi] U > u_{s_0} \, \text{ for all $s_0 \in I$.}
$
\end{lemma}
\begin{proof}{Proof.}
Let $s_0 \in I$.
For a play starting from~$s_0$ denote by $f^{(1)} \in T$ the first visited state in $T \setminus I$.
State~$f^{(1)}$ is undefined if $T \setminus I$ is not visited.
We have
\begin{align*}
        \inf_\pi \, \expectation[s_0,\sigma,\pi] U 
 \ =   \ &\inf_\pi \, \sum_{f \in T \setminus I} \probm_{s_0,\sigma,\pi}(f^{(1)} = f) u_f \\
  \ge \ &\inf_\pi \, \sum_{f \in T \setminus I} \Big(
          \probm_{s_0,\sigma,\pi}(f^{(1)} = f \text{ and } \TB \setminus \Ret)\\
        & \quad\quad\quad\quad + \sum_{s \in I} \probm_{s_0,\sigma,\pi}(f^{(1)} = f \text{ and } \Ret_s) u_s \Big)\\
  =   \ &\inf_\pi \, \probm_{s_0,\sigma,\pi}(\TB \setminus \Ret) + \sum_{s \in I} \probm_{s_0,\sigma,\pi}(\Ret_s) u_s\\
  >   \ &u_{s_0}\,.
\end{align*}
The equality on the second line follows from the definitions of $U$ and~$f^{(1)}$.
The inequality on the third line follows from 
a case split whether the play ever returns to $I$ and 
the definitions of $u_f, \Vis, \Vis_s, \Ret$ and~$\Ret_s$; cf.~\eqref{eq:def-uf}.
The equality on the fourth line follows from the fact that $\TB$ and all $\Ret_s$ require a visit of $T \setminus I$,
and the inequality on the last line follows from \cref{lem:Bu-key-aux}.
\hfill\Halmos
\end{proof}

We will need the following lemma on the existence of $\eps$-optimal
memoryless Max strategies for reachability objectives.
It is a slight extension of previous results; see
\cite[Section 7.7]{MaitraSudderth:DiscreteGambling}, and
\cite[Theorem 12.1]{Flesch-Predtetchinski-Sudderth:2020}. 
Recall that the objective $\reachn{Q}{\top}$ is satisfied by plays that
stay inside the set of states $Q$ until they reach the target state $\top$.

\begin{restatable}[{\cite[Lemma 5]{KMSTW:DGA}}]{lemma}{lemreachuni}\label{lem:reach-uni}
Consider a concurrent game with state space~$S$, where Max may have infinite action sets but Min has only finite action sets.
Let $S_0 \subseteq S$ be a finite set of states, and $\top \in S$ a target state.
For all $\eps > 0$ there are a memoryless Max strategy~$\sigma_0$ and
a finite set $Q$ with $S_0 \subseteq Q \subseteq S$ such that
$\inf_\pi \, \probm_{s,\sigma_0,\pi}(\reachn{Q}{\top}) \ge \valueof{\reach[\top]}{s} - \eps$ holds for all $s \in S_0$.
If the game is turn-based, there is a deterministic such~$\sigma_0$.
\end{restatable}

\Cref{lem:reach-uni} is used in the proof of the following lemma,
to ``replace'' the general Max strategy~$\sigma$ from \cref{lem:Bu-key-aux,lem:Bu-key-aux-1} by a memoryless one.
For a bridge $\Lambda = (I, L, \sigma_0)$ (where $I$~is as before) and plays starting from~$I$, let $U_\Lambda$~denote the random variable such that $U_\Lambda = u_f$ if $f \in T$ is the state that is first visited upon leaving $I \cup L$, and $U_\Lambda = 0$ if $I \cup L$ is not left or the state first visited upon leaving $I \cup L$ is not in~$T$.

\begin{lemma} \label{lem:Bu-key-aux-2}
There is a bridge $\Lambda = (I, L, \sigma_0)$ (where $I$~is as before) such
that $L \cap T = \emptyset$ and for every Max strategy~$\sigma$ consistent with~$\Lambda$ and for all $s_0 \in I$ we have $\inf_\pi \, \expectation[s_0,\sigma,\pi] U_\Lambda > u_{s_0}$.
If $\game_\bot$~is turn-based then $\sigma_0$~is deterministic.
\end{lemma}
\begin{proof}{Proof.}
We transform the ``weighted reachability'' objective defined by the random
variable~$U$ to a reachability objective
by constructing a new game, $\hat\game$, from~$\game_\bot$ as follows.
Add a fresh state, say~$\top$, and for each $f \in T \setminus I$ redirect
each incoming transition so that with probability~$u_f$ it goes to~$\top$ and
with probability~$1-u_f$ to the losing sink~$\bot$.
(This is possible, since $u_f \in (0,1]$.)

Let $\sigma_1$~be the strategy from \cref{lem:Bu-key-aux-1}; i.e., we have
that
$\inf_\pi \, \expectation[\game_\bot,s,\sigma_1,\pi] U > u_{s}$ for all $s \in I$.
Let $\eps > 0$ be such that $\inf_\pi \, \expectation[\game_\bot,s,\sigma_1,\pi] U - \eps > u_{s}$ holds for all $s \in I$.
By \cref{lem:reach-uni} there are a memoryless (if $\game_\bot$~is turn-based: memoryless deterministic) strategy~$\sigma_0$ and a finite set $Q \subseteq (S \setminus T) \cup I$ such that $\inf_\pi \, \probm_{\hat\game,s,\sigma_0,\pi}(\reachn{Q}{\top}) \ge \valueof{\hat\game,\reach[\top]}{s} - \eps$ holds for all $s \in I$.
Define the bridge $\Lambda = (I,Q \setminus I,\sigma_0)$.
Let $s \in I$ be arbitrary, and let $\sigma$~be an arbitrary strategy consistent with~$\Lambda$.
It suffices to show that $\inf_\pi \, \expectation[\game_\bot,s,\sigma,\pi] U_\Lambda > u_{s}$.
We have
\begin{align*}
\inf_\pi \, \expectation[\game_\bot,s,\sigma,\pi] U_\Lambda \
& =   \ \inf_\pi \, \probm_{\hat\game,s,\sigma,\pi}(\reachn{Q}{\top}) \\
& =   \ \inf_\pi \, \probm_{\hat\game,s,\sigma_0,\pi}(\reachn{Q}{\top}) \\
& \ge \ \valueof{\hat\game,\reach[\top]}{s} - \eps \\
& \ge \ \inf_\pi \, \probm_{\hat\game,s,\sigma_1,\pi}(\reach[\top]) - \eps \\
& =   \ \inf_\pi \, \expectation[\game_\bot,s,\sigma_1,\pi] U - \eps \\
& >   \ u_{s}\,,
\end{align*}
as required.
\hfill\Halmos
\end{proof}

Let $\Lambda = (I, L, \sigma_0)$ be the bridge from \cref{lem:Bu-key-aux-2}.
For plays starting from~$I$, define~$\bev{\Lambda}_1$ as the following event:
leave $I \cup L$; upon leaving $I \cup L$ immediately visit a state in $T$.
Plays from~$I$ that are in $\bev{\Lambda}$ are also in~$\bev{\Lambda}_1$
(but not vice-versa, since $\bev{\Lambda}$ requires this to happen after
\emph{each} visit to $I$, and not just once as for $\bev{\Lambda}_1$).

The following lemma strengthens \cref{lem:Bu-key-aux}.

\begin{lemma} \label{lem:Bu-key-item-2}
There is a Max strategy~$\sigma$ consistent with~$\Lambda$ such that,
for all $s_0 \in I$,
\[
 \inf_\pi \, \Big( \probm_{s_0,\sigma,\pi}(\bev{\Lambda} \cap \TB \setminus \Ret) + \sum_{s \in I} \probm_{s_0,\sigma,\pi}(\bev{\Lambda}_1 \cap \Ret_s) u_s \Big) \ > \ u_{s_0} 
\]
\end{lemma}
\begin{proof}{Proof.}
  By \cref{lem:Bu-key-aux-2} there is $x \in (0,1)$ such that for all~$\sigma$
  consistent with~$\Lambda$ and all $s_0 \in I$ we have $x \inf_\pi \, \expectation[s_0,\sigma,\pi] U_\Lambda > u_{s_0}$.
  By the definition of~$u_f$ for $f \in T$ (\Cref{eq:def-uf}),
  there is a Max strategy~$\sigma_2$ such that,
for all $f \in T \setminus I$,
\[
 \inf_\pi \, \Big( \probm_{f,\sigma_2,\pi}(\TB \setminus \Vis) + \sum_{s \in I} \probm_{f,\sigma_2,\pi}(\Vis_s) u_s \Big) \ \ge \ x\, u_f
\]
We can choose~$\sigma_2$ to be consistent with~$\Lambda$.
Let $s_0 \in I$ be arbitrary.
For a play starting from~$s_0$ denote by $f_\Lambda \in T$ the state that is first visited upon leaving $I \cup L$.
State~$f_\Lambda$ is undefined if $I \cup L$ is not left or the state first visited upon leaving $I \cup L$ is not in~$T$.
Let $\sigma$ be a strategy consistent with~$\Lambda$ which switches to~$\sigma_2$ as soon as $f_\Lambda$~is visited.
We have
\begin{align*}
    & \inf_\pi\, \Big(\probm_{s_0,\sigma,\pi}(\bev{\Lambda} \cap \TB \setminus \Ret) + \sum_{s \in I} \probm_{s_0,\sigma,\pi}(\bev{\Lambda}_1 \cap \Ret_s) u_s \Big) \\
= \ & \inf_\pi\, \Big(\probm_{s_0,\sigma,\pi}(\bev{\Lambda}_1 \cap \TB \setminus \Ret) + \sum_{s \in I} \probm_{s_0,\sigma,\pi}(\bev{\Lambda}_1 \cap \Ret_s) u_s \Big) \\
= \ & \inf_\pi\, \sum_{f \in T \setminus I} \Big(
      \probm_{s_0,\sigma,\pi}(f_\Lambda = f \text{ and } \TB \setminus \Ret) \\
  & \quad\quad\quad\quad + \sum_{s \in I} \probm_{s_0,\sigma,\pi}(f_\Lambda = f \text{ and } \Ret_s) u_s \Big) \\
= \ & \inf_\pi\, \sum_{f \in T \setminus I} \Big(
      \probm_{s_0,\sigma,\pi}(f_\Lambda = f) \cdot \Big(
      \probm_{f,\sigma_2,\pi}(\TB \setminus \Vis) \\
  & \hspace{42mm} + \sum_{s \in I} \probm_{f,\sigma_2,\pi}(\Vis_s) u_s \Big) \Big) \\
\ge \ & x \inf_\pi\, \sum_{f \in T \setminus I} \probm_{s_0,\sigma,\pi}(f_\Lambda = f) u_f \\
=   \ & x \inf_\pi\, \expectation[s_0,\sigma,\pi] U_\Lambda \\
>   \ & u_{s_0}\,,
\end{align*}
as required.
\hfill\Halmos
\end{proof}

Now we can prove \cref{lem:Bu-key}.

\lembukey*
\begin{proof}{Proof of \cref{lem:Bu-key}.}
Choose $q, x \in (0,1)$ such that $q x \ge r$.
Let $u \in (0,1]^{I \cup T}$ and the bridge $\Lambda = (I,L,\sigma_0)$ be as defined before.
Let $\sigma$~be the strategy from \cref{lem:Bu-key-item-2}.
Define a strategy~$\sigma^*$ as follows.
For any initial state~$\tilde{s}$, strategy~$\sigma^*$ first plays a strategy~$\sigma_{\tilde{s}}$ with $\inf_\pi \, \probm_{\tilde{s},\sigma_{\tilde{s}},\pi}(\TB) \ge x \valueof{\TB}{\tilde{s}}$;
if $I$~is entered (which is at once if $\tilde{s} \in I$), strategy~$\sigma^*$ switches to~$\sigma$;
from then on, whenever $\bev{\Lambda}_1$ occurs and then the play returns to~$I$, strategy~$\sigma^*$ restarts~$\sigma$ upon that return.
Since $\sigma$~is consistent with~$\Lambda$, so is~$\sigma^*$.
Let $\tilde{s} \in S$ be an arbitrary initial state, and let $\pi$~be an arbitrary Min strategy.
It suffices to show that $\probm_{\tilde{s},\sigma^*,\pi}(\bev{\Lambda} \cap \TB) \ge r \valueof{\TB}{\tilde{s}}$.

For a play starting from~$\tilde{s}$, we define in the following a sequence of
states, namely $s^{(1)}, s^{(2)}, \ldots \in I$.
This sequence may be finite, in which case $s^{(n)}$~is no longer defined from
some $n$ on. Each defined~$s^{(n)}$ naturally corresponds to a visit of~$I$.
Specifically, denote by $s^{(1)} \in I$ the first visited state in~$I$, and,
for $n \ge 1$, denote by $s^{(n+1)} \in I$ the state for which the play
suffix started from~$s^{(n)}$ satisfies $\bev{\Lambda}_1 \cap \Ret_{s^{(n+1)}}$.
In other words, after the visit of~$s^{(n)}$ the event~$\bev{\Lambda}_1$
occurs and then a state in~$I$ is visited and the first such visited state is~$s^{(n+1)}$.
Note that $\sigma^*$ restarts $\sigma$ exactly upon each visit~$s^{(n)}$.
Denote by~$\tau$ the random variable, taking values in
$\{0,1,2,\ldots\} \cup \{\infty\}$, such that for all $n \in \{0, 1, 2,
\ldots\}$ we have that $s^{(n)}$~is defined if and only if $\tau \ge n$.
Let $\mathcal{F}_0$ be the sigma-algebra generated by the
cylinder set corresponding to the history that consists
only of the initial state~$\tilde{s}$.
For all $n \in \{1,2,\ldots\}$, let $\mathcal{F}_n$ be the
sigma-algebra generated by the cylinder sets corresponding
to the histories $w s \in Z^* \times S$ such that $w$~includes
at most the visits $s^{(1)}, \ldots, s^{(n-1)}$ (i.e., the
state~$s$ at the end of history~$w s$ may be the visit $s^{(n)}$).
For all $n \ge 0$, the event $\tau \ge n$ and the random
states $s^{(1)}, \ldots, s^{(\min\{\tau,n\})}$ are $\mathcal{F}_n$-measurable.

Define random variables $X_0 \eqdef r \valueof{\TB}{\tilde{s}}$ and $X_1, X_2, \ldots$ by
\[
X_n \ \eqdef \ \begin{cases} u_{s^{(n)}}     & \text{if } \tau \ge n \\
                             1               & \text{if } \tau <   n \text{ and } \bev{\Lambda} \cap \TB \\
                             0               & \text{if } \tau <   n \text{ and not } (\bev{\Lambda} \cap \TB) \,.
               \end{cases}
\]
Each $X_n$ is $\mathcal{F}_n$-measurable.
We have
\begin{align*}
 \expectation[\tilde{s},\sigma^*,\pi](X_1 \mid \mathcal{F}_0) 
= \ &\sum_{s \in I} \probm_{\tilde{s},\sigma_{\tilde{s}},\pi}(\Vis_s) \cdot u_s + \probm_{\tilde{s},\sigma_{\tilde{s}},\pi}(\bev{\Lambda} \cap \TB \setminus \Vis) \cdot 1 \\
= \ &\sum_{s \in I} \probm_{\tilde{s},\sigma_{\tilde{s}},\pi}(\Vis_s) u_s + \probm_{\tilde{s},\sigma_{\tilde{s}},\pi}(\TB \setminus \Vis) \\
\ge \ & q \sum_{s \in I} \probm_{\tilde{s},\sigma_{\tilde{s}},\pi}(\Vis_s) \valueof{\TB}{s} + \probm_{\tilde{s},\sigma_{\tilde{s}},\pi}(\TB \setminus \Vis) \\
\ge \ & q \Big( \sum_{s \in I}  \probm_{\tilde{s},\sigma_{\tilde{s}},\pi}(\Vis_s) \valueof{\TB}{s} + \probm_{\tilde{s},\sigma_{\tilde{s}},\pi}(\TB \setminus \Vis) \Big) \\
\ge \ & q \inf_{\pi'} \Big( \sum_{s \in I}  \probm_{\tilde{s},\sigma_{\tilde{s}},\pi'}(\Vis_s) \valueof{\TB}{s} + \probm_{\tilde{s},\sigma_{\tilde{s}},\pi'}(\TB \setminus \Vis) \Big) \\
\ge \ & q \inf_{\pi'} \probm_{\tilde{s},\sigma_{\tilde{s}},\pi'}(\TB) \\
\ge \ & q x \valueof{\TB}{\tilde{s}} \\
\ge \ & r \valueof{\TB}{\tilde{s}} \\
 =  \ & X_0\,.
\end{align*}

For all $n \ge 1$ and $s_0 \in I$, write $C_{n,s_0} \subseteq Z^* \times \{s_0\}$ for the set of histories that include the visits $s^{(1)}, \ldots, s^{(n)} \in I$ and end with $s^{(n)} = s_0$.
For any $h \in C_{n,s_0}$, write $\Cyl{h} \subseteq Z^\omega$ for the cylinder set corresponding to~$h$.
For all $n \ge 1$ we have $\{\tau \ge n\} = \bigcup_{s_0 \in I} \bigcup_{h \in C_{n,s_0}} \Cyl{h}$, where the union is disjoint.
For any history~$h$ write $\pi_h$ for the Min strategy whose behavior is exactly like the behavior of~$\pi$ after~$h$.
By the definition of~$\sigma^*$ and by \cref{lem:Bu-key-item-2} we have for all $n \ge 1$ and $s_0 \in I$ and $h \in C_{n,s_0}$
\begin{equation} \label{eq:lem-Bu-key-aux-1-sub}
\begin{aligned}
 \expectation[\tilde s,\sigma^*,\pi](X_{n+1} \mid \Cyl{h}) 
&=\ \probm_{\tilde s,\sigma^*,\pi}(\tau = n \text{ and } \bev{\Lambda} \cap \TB \mid \Cyl{h}) \\
& \hspace{5mm} \mbox{} + \sum_{s \in I} \probm_{\tilde s,\sigma^*,\pi}(\tau \ge n+1 \text{ and } s^{(n+1)}=s \mid \Cyl{h}) u_s \\
&=\ \probm_{s_0,\sigma,\pi_h}(\bev{\Lambda} \cap \TB \setminus \Ret) + \sum_{s \in I} \probm_{s_0,\sigma,\pi_h}(\bev{\Lambda}_1 \cap \Ret_s) u_s \\
&>\ u_{s_0}\,. 
\end{aligned}
\end{equation}
To avoid clutter, let us drop the subscripts from $\probm_{\tilde{s},\sigma^*,\pi}$ and $\expectation[\tilde{s},\sigma^*,\pi]$ in the following.
For all $n \ge 1$
\begin{align*}
 \expectation(X_{n+1} \mid \mathcal{F}_n) \
  &= \ \mathds{1}_{\tau < n} X_n + \sum_{s_0 \in I} \sum_{h \in C_{n,s_0}} \mathds{1}_{\Cyl{h}} \expectation(X_{n+1} \mid \Cyl{h}) \\
  &\mathop{\ge}^{\text{\eqref{eq:lem-Bu-key-aux-1-sub}}} \ \mathds{1}_{\tau < n} X_n + \sum_{s_0 \in I} \sum_{h \in C_{n,s_0}} \mathds{1}_{\Cyl{h}} u_{s_0} \\
  &=\ \mathds{1}_{\tau < n} X_n + \sum_{s_0 \in I} \mathds{1}_{\tau \ge n,\ s^{(n)}=s_0} u_{s_0} \\
  &=\ \mathds{1}_{\tau < n} X_n + \mathds{1}_{\tau \ge n} u_{s^{(n)}} \\
  &=\ \mathds{1}_{\tau < n} X_n + \mathds{1}_{\tau \ge n} X_n \\
  &= \ X_n\,,
\end{align*}
where $\mathds{1}_E$ is the indicator function of~$E$.
That is, $X_0, X_1, X_2, \ldots$ is a submartingale with respect to the filtration $\mathcal{F}_0, \mathcal{F}_1, \mathcal{F}_2, \ldots$.
Random variable $\tau+1$ is a stopping time with respect to the same filtration.
Using the optional stopping theorem
(noting that $|X_n| \le \max\{1,\max_{s \in I} u_s\} \le 1$ for all~$n$), we obtain
\[
r \valueof{\TB}{\tilde{s}} \ = \ \expectation X_0 \ \le \ \expectation X_{\tau+1} \ \le \ \probm(X_{\tau+1} \ne 0) \,,
\]
as $0 \le X_{\tau+1} \le 1$.
It follows from the definition of the random variables~$X_n$ that,
if $\tau < \infty$ and not~$(\bev{\Lambda} \cap \TB)$,
then $X_{\tau+1} = 0$.
Thus,
\begin{align*}
 r \valueof{\TB}{\tilde{s}} \ &\le \ \probm(X_{\tau+1} \ne 0) \ \le \ \probm(\tau=\infty \text{ or } (\bev{\Lambda} \cap \TB))
\\
& \le \ \probm(\tau=\infty) + \probm(\bev{\Lambda} \cap \TB)\,.
\end{align*}
By the construction of~$\game_\bot$ and the strategy~$\sigma^*$,
there is $\delta > 0$ such that upon every visit of~$I$ corresponding to $s^{(1)}, s^{(2)},
\ldots \in I$ the probability of falling into~$\bot$ in the very next step is
at least~$\delta$.
(This is because $\sigma^*$ plays a fixed memoryless strategy upon entering $I$.)
Thus, $\probm(\tau=\infty) = 0$.
Hence, $r \valueof{\TB}{\tilde{s}} \le \probm(\bev{\Lambda} \cap \TB)$, as required.
\hfill\Halmos
\end{proof}

Now we prove the main theorem.

\begin{proof}{Proof of \cref{thm:TrBuchi}}
Let $r \in (0,1)$ be arbitrary.
It is convenient to assume that $S = \{2,4,6,\ldots\}$.
We define a sequence  $(\game_i)_{i\geq 1}$ of games inductively.
Each game~$\game_i$ will have a state space 
\[S_i \eqdef S \cup L'_1 \cup \cdots \cup L'_i\,,\]
where $S, L'_1, L'_2, \ldots$ are pairwise disjoint and $L'_i \subseteq \{3,5,7,\ldots\}$.
We will later associate the even and odd state numbers in these derived games
with different memory modes $0$ and $1$ of a 1-bit strategy in the original game $\game_\bot$, respectively.
 
\begin{figure*}[pt]
\begin{center}
\includegraphics{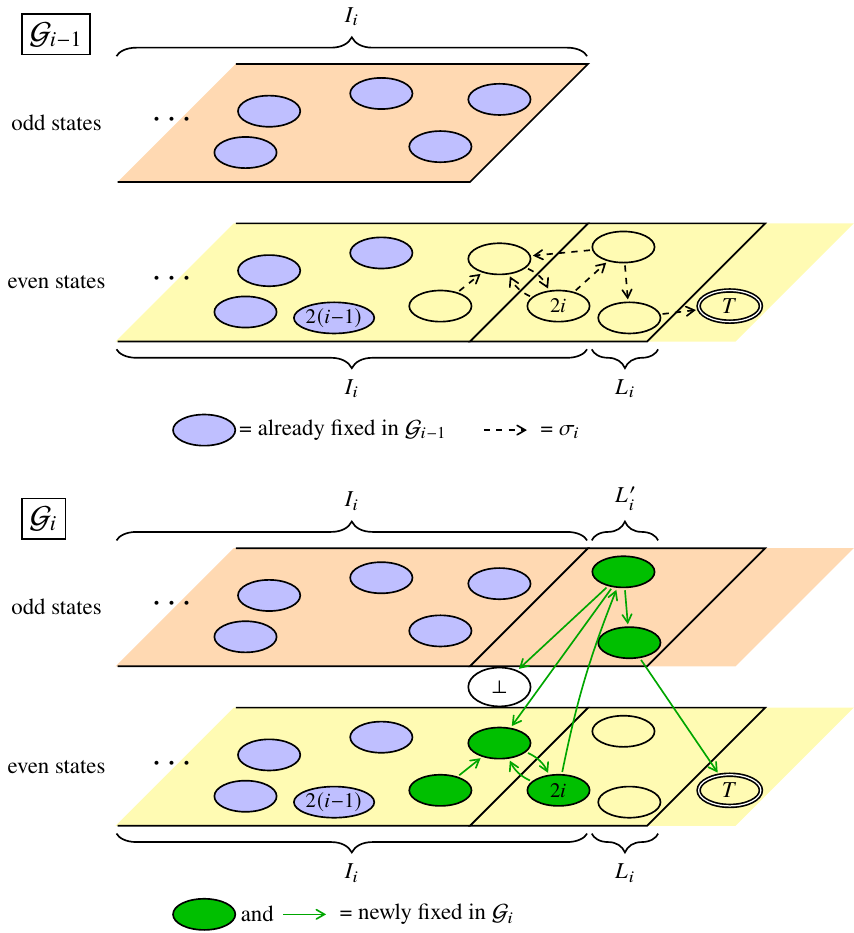}
\end{center}	
\caption{Illustration of the construction in \cref{thm:TrBuchi}.
The top picture indicates a bridge $\Lambda_i = (I_i,L_i,\sigma_i)$ in game~$\game_{i-1}$.
The bottom picture indicates how in game~$\game_i$ the bridge~$\Lambda_i$ is ``plastered'' using~$L'_i$ instead of~$L_i$.
}
\label{fig:bridge}
\end{figure*}
Define $\game_0 \eqdef \game_\bot$ and 
\[I_0 \eqdef L_0 \eqdef L'_0 \eqdef \emptyset\,.\]
For all $i \ge 1$, 
define 
\[I_i \eqdef I_{i-1} \cup L_{i-1} \cup L'_{i-1} \cup \{2 i\}\,.\]
By \cref{lem:Bu-key}, in~$\game_{i-1}$ there are a bridge $\Lambda_i =
(I_i,L_i,\sigma_i)$ and a Max
strategy~$\sigma$ consistent with~$\Lambda_i$ so that
for every $s \in S_{i-1}$ 
we have 
\[\inf_\pi\probm_{\game_{i-1},s,\sigma,\pi}(\bev{\Lambda_i} \cap \TB) \ge r^{2^{-i}} \valueof{\game_{i-1},\TB}{s}\,.\]
If the game is turn-based then the bridge $\Lambda_i$ can be made deterministic.
Since $I_i \cap L_i = \emptyset$ and 
\[I_i \supseteq L'_1 \cup \ldots \cup L'_{i-1}\,,\] all states in~$L_i$ are in~$S$, i.e., are ``even''.
See the top of \cref{fig:bridge}.

We construct~$\game_i$ from~$\game_{i-1}$ by adding fresh, ``odd'', copies of the states in~$L_i$, namely 
\[L'_i \eqdef \{2 j + 1 \mid 2 j \in L_i\}\,.\]
In~$\game_i$ we ``plaster''
the bridge~$\Lambda_i$, but using~$L'_i$ instead
of~$L_i$. (In MDPs/games, ``plastering'' means fixing the choices of a
player in a subset of the states in a certain defined way; here according to the bridge.
See \cite{Ornstein:AMS1969} for an early use of this technique.)
This is to enforce that upon every entry into~$I_i$, a version of~$\sigma_i$
is played by Max until $I_i \cup L'_i$~is left, at which point either a state in~$T$ is visited or the play falls into~$\bot$.
More precisely, we replace the Max actions of states $s \in I_i$ by the single action~$\sigma_i(s)$, and states $2 j + 1 \in L'_i$ get the single action~$\sigma_i(2 j)$.
The successor state distributions of those actions are modified so that transitions to states $2 j \in L_i$ are redirected to the ``odd sibling'' $2 j + 1 \in L'_i$, transitions to~$I_i \cup (T \setminus L_i)$ are kept, and all other transitions are redirected to~$\bot$.
The states in~$L_i$ remain untouched for now (but will be plastered in~$\game_{i+1}$).
See the bottom of \cref{fig:bridge}.

From the properties of the bridge~$\Lambda_i$, for all $s \in S_{i-1}$,  we have 
\begin{align} \label{eq:Bu-val-iter}
 \valueof{\game_i,\TB}{s} \ \ge  \ \inf_\pi
                                  \probm_{\game_{i-1},s,\sigma,\pi}(\bev{\Lambda_i}
                                  \cap \TB)  
  \ge  \ r^{2^{-i}} \valueof{\game_{i-1},\TB}{s}\,.
\end{align}

There is a natural limit of the games~$\game_i$, say $\game_\infty$,
with state space 
\[S_\infty \eqdef S \cup \bigcup_{i \ge 1} L'_i\,,\]
where all choices of Max have been fixed.
(One does not need to fix actions in $\bot$, since its value is zero anyway.)

The game $\game_\infty$ can be seen as an MDP where only Min is still active.
From the point of view of Max, the fixed choices of Max in
$\game_\infty$ define a two-mode (public memory with deterministic updates)
Max strategy, say $\sigma$, for~$\game_\bot$,
where 
\begin{enumerate}
	\item each (even) state $s \in S$ in~$\game_\infty$ corresponds to state~$s$
in~$\game_\bot$ 
and $\sigma$~being in memory mode~$0$,
\item and each (odd) state $s \in \bigcup_{i \ge 1} L'_i$ in~$\game_\infty$
corresponds to state~$s-1$ in~$\game_\bot$ and $\sigma$~being in memory mode~$1$.
\item The start mode is~$0$.
\item The mode update corresponding to transitions that were redirected to~$\bot$
in the construction of~$\game_\infty$ can be defined arbitrarily, say to mode~$0$.
\end{enumerate}

Note that, while the actions of $\sigma$ can be mixed in general, its memory
updates are always deterministic, switching the memory mode after visiting $I_i$ and
$T$, respectively.
However, if the game is turn-based then all the bridges can be made deterministic,
and thus $\sigma$ is completely deterministic.

Recall that $\sigma[0]$ refers to $\sigma$ starting in memory mode $0$.
It remains to show that $\sigma[0]$ 
is nearly optimal from every state, that is, for all $s \in S$
\[\inf_\pi\probm_{\game_\bot,s,\sigma[0],\pi}(\TB) \ge r \valueof{\game_\bot}{s}.\]

\smallskip
Recall that $\game_\infty$ is an MDP where only Min plays.
For all $s \in S$, the following holds:
\[\inf_\pi \probm_{\game_\bot,s,\sigma[0],\pi}(\TB) \ge \inf_\pi\probm_{\game_\infty,s,\pi}(\TB)\,.\]
We note in passing that
the inequality above might be strict due to the transitions redirected to~$\bot$ in~$\game_\infty$.
The game $\game_\infty$ has been obtained from $\game_\bot$ by fixing a
finite-memory (actually 1-bit) Max strategy. Thus, by our construction of
$\game_\bot$, the events $\transient$ and~$\avoid[\bot]$ are equal up to measure zero
in $\game_\infty$, under any Min strategy (by \Cref{lem:strongerTrNotbot}).
In~$\game_\infty$, for any state $i \in S_\infty$, the only way out of the finite set $I_i \cup L'_i$ is via a state in~$T$ or falling into~$\bot$.
Thus, in~$\game_\infty$ the events $\Bu$ and~$\transient$ are also equal up to
measure zero under any Min strategy.
Hence, for all $s \in S$ we have
\begin{equation} \label{eq:Bu-chain-first-part}
  \inf_\pi \probm_{\game_\bot,s,\sigma[0],\pi}(\TB) \ \ge 
\inf_\pi\probm_{\game_\infty,s,\pi}(\TB) \\
=  \inf_\pi\probm_{\game_\infty,s,\pi}(\transient) \\ 
=  \inf_\pi\probm_{\game_\infty,s,\pi}(\avoid[\bot])\,.
\end{equation}
The rest of the proof proceeds similarly to \Cref{sec-trans}.
Let us write $v_i(s) \eqdef \valueof{\game_i,\TB}{s}$ in the following to avoid clutter.
For $i \ge 0$ define $r_i \eqdef \prod_{j>i} r^{2^{-j}}$.
It follows from~\eqref{eq:Bu-val-iter} that for all $i \ge 1$ and all $s \in S_{i-1}$ 
\begin{equation} \label{eq:Bu-r-value-inv}
 r_{i} v_i(s) \ \ge \ r_{i-1} v_{i-1}(s) \,.
\end{equation}

Recall that $H_{n-1}$~is the set of histories with the first $n$~states.
For all $n \ge 1$, let $\mathcal{F}_n$ be the sigma-algebra generated by the cylinder sets corresponding to~$H_{n-1}$.
For a play visiting states $s_1, s_2, \ldots$ and $n \ge 1$ define \[m(n) \eqdef \max\{s_1, \ldots, s_n\}\] and the random variable~$Y_n$, taking values in~$[0,1]$, by 
\[Y_n \eqdef r_{m(n)} v_{m(n)}(s_n)\,.\]
Both $m(n)$ and~$Y_n$ are $\mathcal{F}_n$-measurable.
Intuitively, $Y_n$~is obtained by scaling down $v_{m(n)} = \valueof{\game_{m(n)}}{s_n}$, which is the value in~$\game_{m(n)}$ where (at least) the states $s_1, \ldots, s_n$ have been plastered, by a lower bound, $r_{m(n)}$, on the proportion of the value that will be maintained by ``future'' plasterings.

Let $s_1 \in S$ be an arbitrary state.
It suffices to show that \[\inf_\pi\probm_{\game_\bot,s_1,\sigma[0],\pi}(\TB) \ge r \valueof{\game_\bot}{s_1}\,.\]

The following claim follows from the fact that $(Y_n)_{n=1}^{\infty}$ is a submartingale
with respect to the filtration $(\mathcal{F}_n)_{n=1}^{\infty}$.

\begin{restatable}{claim}{eqBusubmart}\label{eq:Bu-submart}
For any Min strategy $\pi$,
$
\expectation[\game_\infty,s_1,\pi] Y_{n} \ \ge \ \expectation[\game_\infty,s_1,\pi] Y_{1} \quad\text{for all $n \ge 1$}.
$
\end{restatable}
\begin{proof}{Proof of the claim.}
For all $n \ge 1$ and any Min strategy $\pi$ we have
\begin{align*}
\expectation[\game_\infty,s_1,\pi](Y_{n+1} \mid \mathcal{F}_n) \
&   = \ \expectation[\game_\infty,s_1,\pi](r_{m(n+1)} v_{m(n+1)}(s_{n+1}) \mid \mathcal{F}_n) \\
& \ge \ \expectation[\game_\infty,s_1,\pi](r_{m(n)} v_{m(n)}(s_{n+1}) \mid \mathcal{F}_n) && \text{by~\eqref{eq:Bu-r-value-inv}} \\
&  \ge \ \expectation[\game_\infty,s_1,\pi](r_{m(n)} v_{m(n)}(s_n) \mid \mathcal{F}_n) \ = \ Y_n && \text{see below,}
\end{align*}
where the inequality marked with ``see below'' follows from the
observation that the successor state distribution in state~$s_n$
of~$\game_\infty$ has been fixed already in~$\game_{m(n)}$,
and, therefore, the expected value, in terms of~$\game_{\infty}$,
cannot drop in one step (though it might increase if $\pi$ plays poorly).
Thus, $Y_1, Y_2, \ldots$ is a submartingale
w.r.t.\ $\expectation[\game_\infty,s_1,\pi]$ for any Min
strategy $\pi$ and thus the claim follows.
\hfill\Halmos
\end{proof}

Hence, 
\begin{align*}
& \inf_\pi \probm_{\game_\bot,s_1,\sigma[0],\pi}(\TB) 
\ge \ \inf_\pi\probm_{\game_\infty,s_1,\pi}(\avoid[\bot]) && \text{by~\eqref{eq:Bu-chain-first-part}} \\
\intertext{Since $\bot$ is a sink state and  has value~$0$,  we have}
& \ge \ \inf_\pi\probm_{\game_\infty,s_1,\pi}\Big(\limsup_{n \to \infty} Y_n > 0\Big) &&  \\
\intertext{Furthermore, since \text{$Y_n \le 1$}, it follows that}
& \ge \ \inf_\pi\expectation[\game_\infty,s_1,\pi]\Big(\limsup_{n \to \infty} Y_n\Big) &&  \\
& \ge \ \inf_\pi\limsup_{n \to \infty}~ \expectation[\game_\infty,s_1,\pi] Y_n && \text{reverse Fatou lemma} \\
& \ge \ \inf_\pi\expectation[\game_\infty,s_1,\pi] Y_{1} && \text{by~\Cref{eq:Bu-submart}} \\
&   = \ r_{s_1} v_{s_1}(s_1) && \text{definitions of $Y_1, m(1)$} \\
& \ge \ r_0 v_0(s_1) && \text{by~\eqref{eq:Bu-r-value-inv}} \\
&   = \ r \valueof{\game_\bot}{s_1} && \text{definitions of~$r_0, \game_0$\,,}
\end{align*}
as required.
This concludes the proof of Theorem~\ref{thm:TrBuchi}.
\hfill\Halmos
\end{proof}

\subsection{Proof of \Cref{thm:Buchi}} \label{sub:proof-thm-Buchi}

In this subsection we prove the following theorem.

\thmBuchi*

In Section~\ref{sec-overview} we claimed that Theorem~\ref{thm:Buchi} is implied by Theorem~\ref{thm:TrBuchi}.
This claim follows from \cref{prop:redacyclic} below.
The proof follows an idea from \cite[Lemma 4]{KMST2020c}:
to show the existence of  1-bit Markov strategies in B\"uchi games, it suffices to
show that  1-bit strategies suffice in a B\"uchi game that is made acyclic
by encoding a step counter in the state space. 

\begin{proposition}
\label{prop:redacyclic}
Suppose that for every \emph{acyclic} B\"uchi game~$\game'$ and $\eps>0$
Max has a 1-bit strategy $\sigma'$ such that
\begin{enumerate}
    \item $\zstrat'[0]$ is multiplicatively $\eps$-optimal from every state,
    \item all memory updates $\stratUp'(\cdot)$ are Dirac, and
    \item $\sigma'$ is deterministic if $\game$~is turn-based.
\end{enumerate}
Then for every B\"uchi game~$\game$ and $\eps>0$
Max has a 1-bit Markov strategy $\sigma$ such that
\begin{enumerate}
    \item $\zstrat[0]$ is multiplicatively $\eps$-optimal from every state,
    \item all memory updates $\stratUp(\cdot)$ are Dirac, and
    \item $\sigma$ is deterministic if $\game$~is turn-based.
\end{enumerate}
\end{proposition}

\begin{proof}{Proof of Proposition~\ref{prop:redacyclic}}
Let $\game$ be a B\"uchi game with set~$S$ of states, where at each state $s\in S$ Max and Min have action sets $A(s)$ and $B(s)$, respectively.
Let $p$ be the transition function of~$\game$.
Let $\eps > 0$.
	
We transform $\game$ into an acyclic B\"uchi game $\game'$ by encoding a step counter into the states. That is, the set of states in  $\game'$ is $S \times \nat$. For each state~$(s,k)$, with $k\in \nat$, Max and Min have action $A(s)$ and $B(s)$, respectively. 
The  transition function~$p'$ of $\game'$ is defined by 
\[p'((s,k),a,b)(s',j)\begin{cases}
		p(s,a,b)(s') & \text{if } j=k+1\\
		0 & \text{otherwise.} 
\end{cases} \]
Note that for each state~$s \in S$ and each $k \in \nat$ the value of~$s$ in~$\game$ equals the value of $(s,k)$ in~$\game'$.
	
By the assumption of the lemma statement, in~$\game'$ Max has a 1-bit strategy~$\sigma'$ with the properties (1)--(3) from the assumption.
Obtain from~$\sigma'$ the 1-bit Markov strategy~$\sigma$, which, at state~$s$ with memory mode $b \in \{0,1\}$ and step counter~$n$, plays like $\sigma'$ at state $(s,n)$ with mode~$b$.
Then $\sigma$ satisfies the properties (1)--(3) from the conclusion of the lemma statement.
In particular, since $\sigma'[0]$ is $\eps$-optimal from every state in~$\game'$, it follows that $\sigma[0]$ is $\eps$-optimal from every state in~$\game$.
Indeed, $\sigma[0]$ is $\eps$-optimal from every state in~$\game$ and for every value of the step counter, thus specifically also for the initial step counter value of~$0$.
\hfill\Halmos
\end{proof}

\section{Strategy Complexity for Minimizer}
\label{sec-minimizer}
We have shown that $\eps$-optimal Max strategies in countable concurrent
B\"uchi games with finite Min action sets
require just a step counter plus 1 bit of public memory
(\Cref{thm:Buchi}), but cannot use less
memory in general, even if the game is finite (\Cref{sec-lower-bounds}).

This upper bound does not hold for the strategy complexity of the opposing Min player.
From player Min's point of view, the objective is to maximize the probability
of satisfying the dual \emph{co-B\"uchi}
objective of visiting the set of target states only \emph{finitely} often.
I.e., Min in a B\"uchi game is the same as Max in a co-B\"uchi game.
However, a step counter plus arbitrary finite (private) memory are \emph{not}
sufficient for $\eps$-optimal Max strategies for the co-B\"uchi objective
in countable concurrent games. In fact, this lower bound holds even for
countable finitely branching \emph{turn-based} co-B\"uchi games \cite[Remark 1]{KMSTW:DGA}.
On the other hand, in \emph{finite-state} concurrent games,
$\eps$-optimal Max strategies for the co-B\"uchi objective can be chosen memoryless randomized
\cite{Alfaro-Henzinger:LICS2000}, and this holds even for the
more general $\liminf$ objective \cite{Secchi:1998}.
\emph{Optimal} Max strategies for the co-B\"uchi objective
need not exist, but might require memory if they do.
\citeauthor{bordais_et_al:LIPIcs.FSTTCS.2022.33}\cite[Section 6]{bordais_et_al:LIPIcs.FSTTCS.2022.33}
give an example of a finite-state co-B\"uchi game where optimal Max strategies
exist but require infinite memory.

For the Transience objective, $\eps$-optimal Max strategies can be chosen memoryless (\Cref{thm:Tr}),
but the strategy complexity is much higher for the opposing Min player.
From player Min's point of view, the objective is to maximize the probability
of satisfying the dual \emph{Recurrence}
objective $\overline{\transient}$ of visiting some state infinitely often.
I.e., Min in a Transience game is the same as Max in a Recurrence game.
The main problem is that, in a Recurrence game with an infinite number of
states, Max cannot commit to visiting
any \emph{particular} state (or any particular finite subset of the states) infinitely
often, because that might be under the opposing player's control.
(Recurrence holds trivially if the game has only finitely many states.)
Even in countably infinite finitely branching turn-based Recurrence games,
Max strategies that use just a step counter plus arbitrary finite (private)
memory are worthless. This can be shown along similar lines as for
turn-based finitely branching co-B\"uchi
games in \cite[Remark 1]{KMSTW:DGA}.

\tikzset{MDPrand/.style={draw,circle,minimum size=11*1.5,inner sep=0}}
\tikzset{MDPcont/.style={draw,rectangle,minimum size=9*1.5,inner sep=0}}

\begin{figure*}[tbp]
 \centering
 \includegraphics{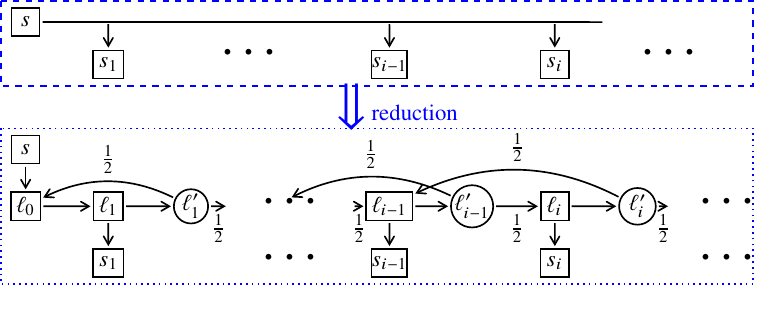} 
\caption{A reduction of infinite Min branching to finite Min branching, provided
  that Min is trying to avoid recurrence.
(Adapted from \cite[Figure 3]{KMST:CONCUR2021}.)}\label{fig:gamblers-ruin-ladder}
\end{figure*}

\begin{restatable}{theorem}{thmRecurrencehard}\label{thm:Recurrence-hard}
There exists a countably infinite finitely branching turn-based game $\game$ with initial
state $\state_0$ with the recurrence objective $\overline{\transient}$ such
that Max can win almost surely from $\state_0$, but
every Max strategy $\zstrat$ that uses only a step counter plus arbitrary finite private
memory is worthless, i.e.,
$\inf_{\ostrat \in \ostratset}\probm_{\game,\state_0,\zstrat,\ostrat}(\overline{\transient})=0$.
\end{restatable}
\begin{proof}{Proof.}
By \cite[Theorem 15]{KMSTW:DGA}, there exists a countably infinite turn-based
stochastic game $\game$, finitely branching for Max but infinitely branching for Min,
such that, for the reachability objective $\reach[\{f\}]$,
Max can win almost surely from $\state_0$,
but every Max strategy $\zstrat$ that uses only a step counter plus arbitrary finite private
memory is worthless, i.e.,
$\inf_{\ostrat \in \ostratset}\probm_{\game,\state_0,\zstrat,\ostrat}(\reach[\{f\}])=0$.

It follows immediately that $\game$ can be modified such that the step counter
from $\state_0$ is encoded into the states, without affecting the properties
above. This modification makes the game acyclic.

We now modify the game further by making the target state $f$ a sink with a
self-loop, and call the new game $\game'$.
In $\game'$ a play is recurrent if and only if it reaches the state $f$, i.e.,
the objectives $\overline{\transient}$ and $\reach[\{f\}]$ coincide.
Hence Max can win $\overline{\transient}$ almost surely from $\state_0$,
but every Max strategy $\zstrat$ that uses only a step counter plus arbitrary finite private
memory is worthless, i.e.,
$\inf_{\ostrat \in \ostratset}\probm_{\game',\state_0,\zstrat,\ostrat}(\overline{\transient})=0$.

The only remaining problem is that Min has infinite branching in $\game'$.
We now modify $\game'$ to obtain a finitely branching game $\game''$ with the
same properties.
To this end, we encode every infinite Min branching of the form
$\state \to \state_i$ for all $i \in \N$ into binary branching by using the
gadget depicted in \Cref{fig:gamblers-ruin-ladder}. It closely resembles
the classic Gambler's ruin with restarts and a $1/2$ chance of winning.
Unless Min chooses some step $l_i \to \state_i$, every state $l_i$ will
be visited infinitely often with probability $1$.
This makes it possible for Min to choose any state $\state_i$ as before.
On the other hand, except for a nullset, all plays that don't exit the
gadget are recurrent. So Min is forced to leave the gadget eventually,
or else she concedes the game to Max.

The last obstacle is that the gadget of \Cref{fig:gamblers-ruin-ladder} does
not preserve path lengths. I.e., the step counter in the new game is different,
and thus one cannot immediately carry step-counter-plus-finite-memory strategies between the games.

However, by the proof of \cite[Theorem 15]{KMSTW:DGA}, the only feature
required to negate the usefulness of the step counter for Max is the following.
Immediately before every Max decision, Min gets to choose 
an arbitrary but finite delay.
In \cite[Theorem 15]{KMSTW:DGA}, Min chooses this delay via a gadget with
infinite branching. However, in the context of our recurrence objective,
the same can be encoded much simpler by giving Min an extra state with a
self-loop just before every Max-state. So Min can delay for as long as she
likes, but she cannot stay in such a self-loop forever, because that would be
recurrent and concede the game to Max.
The proof that this modified game $\game''$ has the required properties
is thus almost identical to the proof of \cite[Theorem 15]{KMSTW:DGA}.
\hfill\Halmos
\end{proof}

\ignore{
\paragraph{Generalized transience.}
In countably infinite co-B\"uchi games with an infinite set of target states $\reachset$
there can exist plays that visit $\reachset$ infinitely often, but visit each
state in $\reachset$ only finitely often.
By the standard definition, such plays do not satisfy the co-B\"uchi objective.
One could define a weaker objective, \emph{weak-co-B\"uchi}, that only
requires that each state in $\reachset$ is visited only finitely often.
Weak-co-B\"uchi coincides with co-B\"uchi if $\reachset$ is finite, and thus
in particular in all finite-state games.
However, in infinite-state games, weak-co-B\"uchi could be very different.
In fact, it can be seen as a generalized transience objective, where one
demands transience only wrt.~the states in the set $\reachset$ instead of all states.
We conjecture that $\eps$-optimal Max strategies for this generalized
transience objective could be memoryless, along similar lines as our proofs
in \Cref{sec-trans}.
}

%
\begin{acks}
The authors thank the anonymous referees for helpful comments and James Main for his careful reading of the manuscript and the valuable feedback.
  
Research supported by EPSRC grants EP/V025848/1 and EP/X042596/1
and ANR grant VeSyAM (ANR-22-CE48-0005).
\end{acks}

\bibliographystyle{ACM-Reference-Format}
\bibliography{journals,conferences,refs}

\end{document}